\documentclass{aa}
\usepackage{rotating}
\usepackage{graphicx}
\usepackage{txfonts}
\begin{document}
\def\teff{$T_{\rm eff}$}
\def\logg{$\log g$}
\def\micro{$\xi$ }
\def\kms{km s$^{-1}$}
\def\p{$\pm$}
\def\vsini{$v\sin i$}

\title{Phase resolved X-ray spectroscopy of HDE\,228766: Probing the wind of an extreme Of$^+$/WNLha star\thanks{Based on observations collected with {\it XMM-Newton}, an ESA science mission with instruments and contributions directly funded by ESA member states and the USA (NASA), and on data collected at the San Pedro M\'artir observatory (Mexico).}}
\author{G.\ Rauw\inst{1},  L.\ Mahy\inst{1}, Y.\ Naz\'e\inst{1}\thanks{Research Associate FRS-FNRS (Belgium)}, P.\ Eenens\inst{2}, J.\ Manfroid\inst{1}\thanks{Honorary Research Director FRS-FNRS (Belgium)}, \and C.A.\ Flores\inst{2}}
\offprints{G.\ Rauw}
\mail{rauw@astro.ulg.ac.be}
\institute{Groupe d'Astrophysique des Hautes Energies, Institut d'Astrophysique et de G\'eophysique, Universit\'e de Li\`ege, All\'ee du 6 Ao\^ut, B\^at B5c, 4000 Li\`ege, Belgium
\and Departamento de Astronomia, Universidad de Guanajuato, Apartado 144, 36000 Guanajuato, GTO, Mexico}
\abstract{HDE\,228766 is a very massive binary system hosting a secondary component, which is probably in an intermediate evolutionary stage between an Of supergiant and an WN star. The wind of this star collides with the wind of its O8\,II companion, leading to relatively strong X-ray emission.}{Measuring the orbital variations of the line-of-sight absorption toward the X-ray emission from the wind-wind interaction zone yields information on the wind densities of both stars.}{X-ray spectra have been collected at three key orbital phases to probe the winds of both stars. Optical photometry has been gathered to set constraints on the orbital inclination of the system.}{The X-ray spectra reveal prominent variations of the intervening column density toward the X-ray emission zone, which are in line with the expectations for a wind-wind collision. We use a toy model to set constraints on the stellar wind parameters by attempting to reproduce the observed variations of the relative fluxes and wind optical depths at 1\,keV.}{The lack of strong optical eclipses sets an upper limit of $\sim 68^{\circ}$ on the orbital inclination. The analysis of the variations of the X-ray spectra suggests an inclination in the range 54 -- 61$^{\circ}$ and indicates that the secondary wind momentum ratio exceeds that of the primary by at least a factor 5. Our models further suggest that the bulk of the X-ray emission arises from the innermost region of the wind interaction zone, which is from a region whose outer radius, as measured from the secondary star, lies between 0.5 and 1.5 times the orbital separation.}
\authorrunning{Rauw et al.}
\titlerunning{X-ray observations of HDE\,228766}
\maketitle

\section{Introduction}\label{intro}
Massive binary systems host a wind interaction region, where the stellar winds of the two stars collide with each other. This wind interaction region can produce a variety of observational signatures from synchrotron radio emission, due to relativistic electrons that are accelerated in the hydrodynamical shocks, to phase-locked optical emission line profile variations, due to the overdensity of the material in the interaction zone (for a review, see Rauw \cite{Brno}). However, the most spectacular signature of this phenomenon is expected in the X-ray domain. Indeed, the large wind velocities ahead of the shocks lead to a substantial increase in the plasma temperature behind the shock, and hydrodynamical simulations predict a strong X-ray emission (e.g., Stevens et al.\ \cite{SBP}). Not only are these objects expected to be overluminous in X-rays compared to single stars of the same spectral type, but their X-ray emission is also expected to vary as a function of orbital phase (Pittard \& Parkin \cite{PP}). The latter modulation stems from either the changing orientation of the system (leading to a varying optical depth along the line of sight across the stellar winds) or changes in the intrinsic emission of the shock (for eccentric binary systems). Previous X-ray observations (mainly with {\it EINSTEIN} and {\it ROSAT}) yielded results in qualitative agreement with these expectations (e.g., Corcoran \cite{Corcoran}), but it was only thanks to phase-resolved X-ray observations with {\it XMM-Newton} and {\it Chandra} that a quantitative picture of the X-ray emission for such interacting wind massive binaries arose (e.g., Rauw \cite{Leuven}). 

An interesting system in this context is the 10.7\,day period early-type binary HDE\,228766, which has a circular orbit (Rauw et al.\ \cite{hde}). The evolutionary stage of this system has been a mystery for several decades. First classified as a Wolf-Rayet binary (e.g., Hiltner \cite{Hiltner}), it was later reclassified as an Of-type system (e.g.\ Massey \& Conti \cite{MC}) and, more recently, as consisting of an O7 primary and an evolved secondary in transition between an Of$^+$ star and a WN8ha Wolf-Rayet object (Rauw et al.\ \cite{hde}). Sota et al.\ (\cite{Sota}) alternatively assigned a spectral-type O4\,If + O8\,II by comparing a blue-violet spectrum of HDE\,228766 at a resolving power of 2500 with linear combinations of spectra of O-type standard stars. The classification proposed by Sota et al.\ (\cite{Sota}) is based on the spectral morphology only and does not formally identify either star as the primary or secondary. Their database of standard stars does not include WN stars, and some of the spectral features seen in the spectrum of HDE\,228766 (strength of the N\,{\sc iv} $\lambda$\,4058 emission and N\,{\sc v}\,4603 absorption; lack of C\,{\sc iii} $\lambda\lambda$\,4647 --51 emission) cannot be explained by the combination of spectral types advocated by Sota et al.\ (\cite{Sota}).

In any case, HDE\,228766 clearly hosts an extreme and interesting massive star. In fact, the optical spectrum of the secondary displays lines from three different ionization stages of nitrogen (N\,{\sc iii}, N\,{\sc iv}, and N\,{\sc v}), an unusual feature, which cannot be easily explained with non-LTE model atmosphere codes that do not account for Auger-type ionization (Rauw et al.\ \cite{hde}). Moreover, the secondary star was found to undergo substantial mass loss at a rate of $\sim 10^{-5}$\,M$_{\odot}$\,yr$^{-1}$ (Rauw et al.\ \cite{hde}).

X-ray emission from this system was first measured with the {\it EINSTEIN} satellite (Chlebowski et al.\ \cite{CHS}). A pointed {\it ROSAT} observation at phase 0.64 (i.e., when the primary is moving away from us, where phase 0.0 corresponds to the conjunction with the primary in front) revealed an X-ray overluminosity by a factor 3.6 (Rauw et al.\ \cite{hde}) as compared to the canonical $L_{\rm X}/L_{\rm bol}$ relation of Bergh\"ofer et al.\ (\cite{Berghoefer}). Phase-resolved optical spectroscopy further revealed a variable H$\alpha$ emission: Doppler tomography of this line showed that part of this emission is very likely formed in a wind interaction region between the components of HDE\,228766, where the more energetic wind of the secondary star either encounters the primary wind or hits the surface of the primary (Rauw et al.\ \cite{hde}).

Since HDE\,228766 has a circular orbit, any variation of its X-ray spectrum will be related to the changing column density along the line of sight rather than to a changing separation between the stars. This is a very convenient situation to diagnose the actual wind column density and thus constrain the mass-loss rates of the binary components. 

\section{Observations}
\begin{figure}[h!tb]
\begin{center}
\resizebox{8cm}{!}{\includegraphics{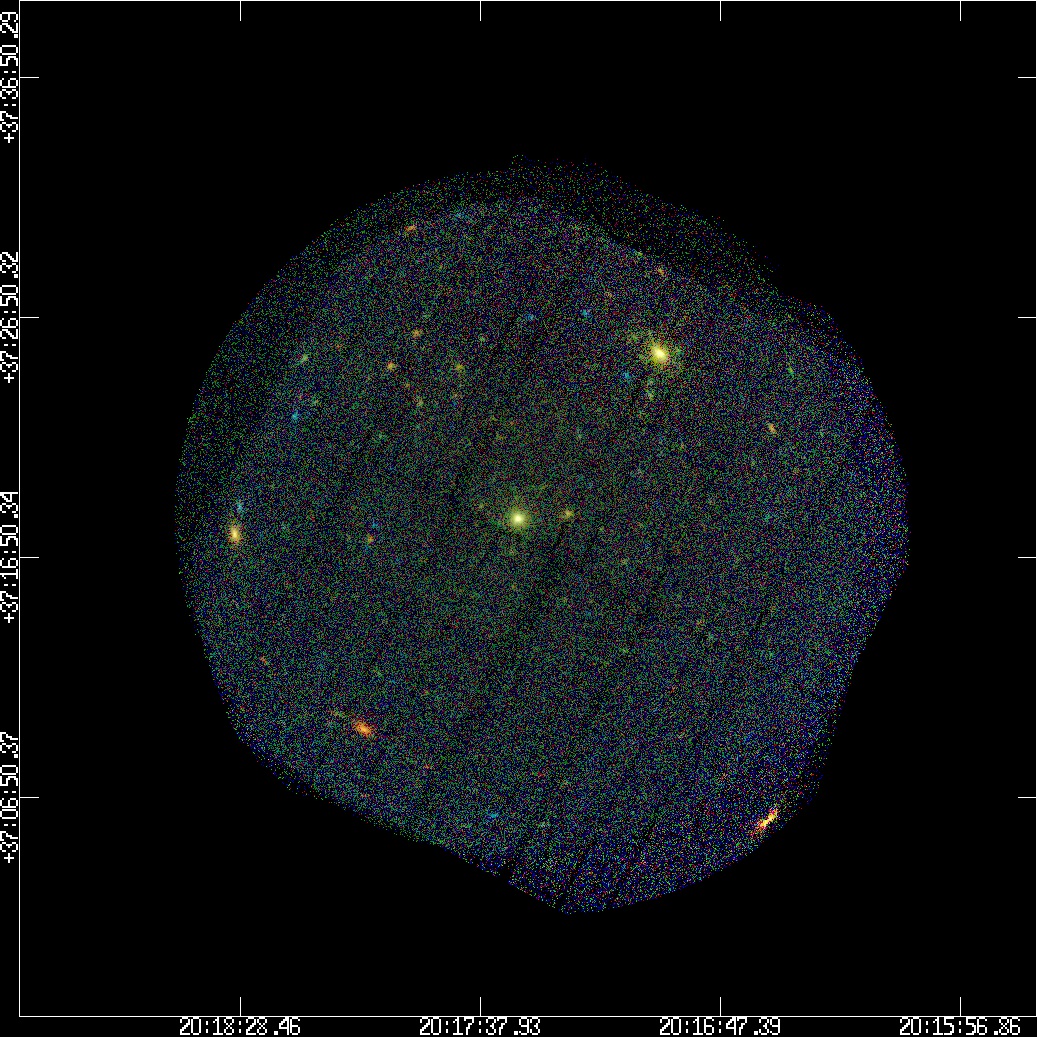}}
\end{center}
\caption{Energy-coded, three-color X-ray image of the field of view around HDE\,228766 built from the EPIC data of the three observations. North is up and east is to the left. The red, green, and blue colors correspond to the photons in the energy ranges of 0.5 -- 1.0, 1.0 -- 2.0, and 2.0 -- 8.0\,keV, respectively. The individual images were exposure corrected before they were combined. The bright source to the northwest of HDE\,228766 is the Wolf-Rayet binary WR\,138 (Palate et al.\ \cite{Palate})\label{colour}.}
\end{figure}

We collected three observations of HDE\,228766 with {\it XMM-Newton} (Jansen et al.\ \cite{Jansen}). These observations were scheduled at specific orbital phases ($\phi = 0.0$, 0.5, and 0.75) during two consecutive binary orbits in May 2011 (ObsIDs 0670480401, 0670480501, and 0670480601, PI G.\ Rauw). Each of these observations had a nominal duration of 22 - 25\,ks. All observations were taken with the EPIC instruments (Turner et al.\ \cite{MOS}, Str\"uder et al.\ \cite{pn}) in full frame mode and using the medium filter to reject optical and UV photons. The raw data were processed with SAS software version 12.0. The first two observations were affected by moderate background flares, so-called soft-proton flares, and we rejected the corresponding time intervals in our processing. An energy-coded image of the field of view is shown in Fig.\,\ref{colour}. The spectra of the source were extracted over a circular region that is centered on the coordinates of HDE\,228766 and adopting an extraction radius of 35\,arcsec. The background spectrum was evaluated from a nearby, source-free circular region located on the same detector chip as the source itself. The prominent X-ray source to the northwest of HDE\,228766 (see Fig.\,\ref{colour}) is the Wolf-Rayet binary WR\,138. The corresponding data were analysed previously by Palate et al.\ (\cite{Palate}).

We also processed the data collected with the RGS reflection spectrometer (den Herder et al.\ \cite{RGS}). However, given the relatively modest count rate and the rather short integration time, the RGS spectra do not provide useful data for HDE\,228766. 

\begin{table}
\caption{Journal of the {\it XMM-Newton} observations.}
\begin{center}
\begin{tabular}{c c c c c}
\hline
Obs & Date          & $\phi$ & \multicolumn{2}{c}{Useful exposure time} \\
    & JD$-$2450000  &        & MOS  & pn \\
    &               &        & (ks) & (ks)  \\
\hline
I   & 5686.703      & $0.750 \pm 0.013$   & 21.6 & 14.1 \\
II  & 5694.768      & $0.501 \pm 0.015$   & 19.9 & 13.0 \\
III & 5700.152      & $0.002 \pm 0.015$   & 23.8 & 18.0 \\
\hline
\end{tabular}
\tablefoot{The date of the observation is given at mid-exposure. The uncertainties on the orbital phase indicate the phase interval corresponding to half of the duration of each observation.}
\end{center}
\end{table}

The Optical Monitor (OM, Mason et al.\ \cite{OM}) aboard {\it XMM-Newton} observed HD\,228766 through the $UVW1$, $UVM2$ and $UVW2$ filters using the `imaging mode'. The data were processed with the relevant commands of the SAS software. 
The detector of the OM has a read-out time of 11\,ms and bright sources suffer from coincidence losses that amount typically to 10\% for sources with 10\,counts\,s$^{-1}$ (Mason et al.\ \cite{OM}). These losses are partially recovered during the data processing to provide `corrected' count rates, which are then converted into AB magnitudes (Oke \cite{Oke}). In the case of HD\,228766, the count rates in the $UVW1$ and $UVM2$ images are so large that the corresponding magnitudes are unreliable. The situation is better for the $UVW2$ filter (effective wavelength 2120\,\AA\ and bandwidth 500\,\AA), where the corrected count rate is around 25\,counts\,s$^{-1}$. 

The $U\,B\,V$ optical photometry of HDE\,228766 was acquired at the San Pedro M\'artir (SPM) Observatory with the 0.84\,m Cassegrain reflector equipped with the photon-counting photometer Cuentapulsos (utilizing an RCA 31034 photomultiplier). A typical observing sequence was sky -- comp1 -- sky -- comp2 -- sky -- HDE\,228766, where comp1 and comp2 stand for the comparison stars HD\,188892 (B5\,IV) and HD\,193369 (A2\,V), respectively.

\begin{figure*}[t!hb]
\begin{minipage}{8cm}
\begin{center}
\resizebox{9cm}{!}{\includegraphics{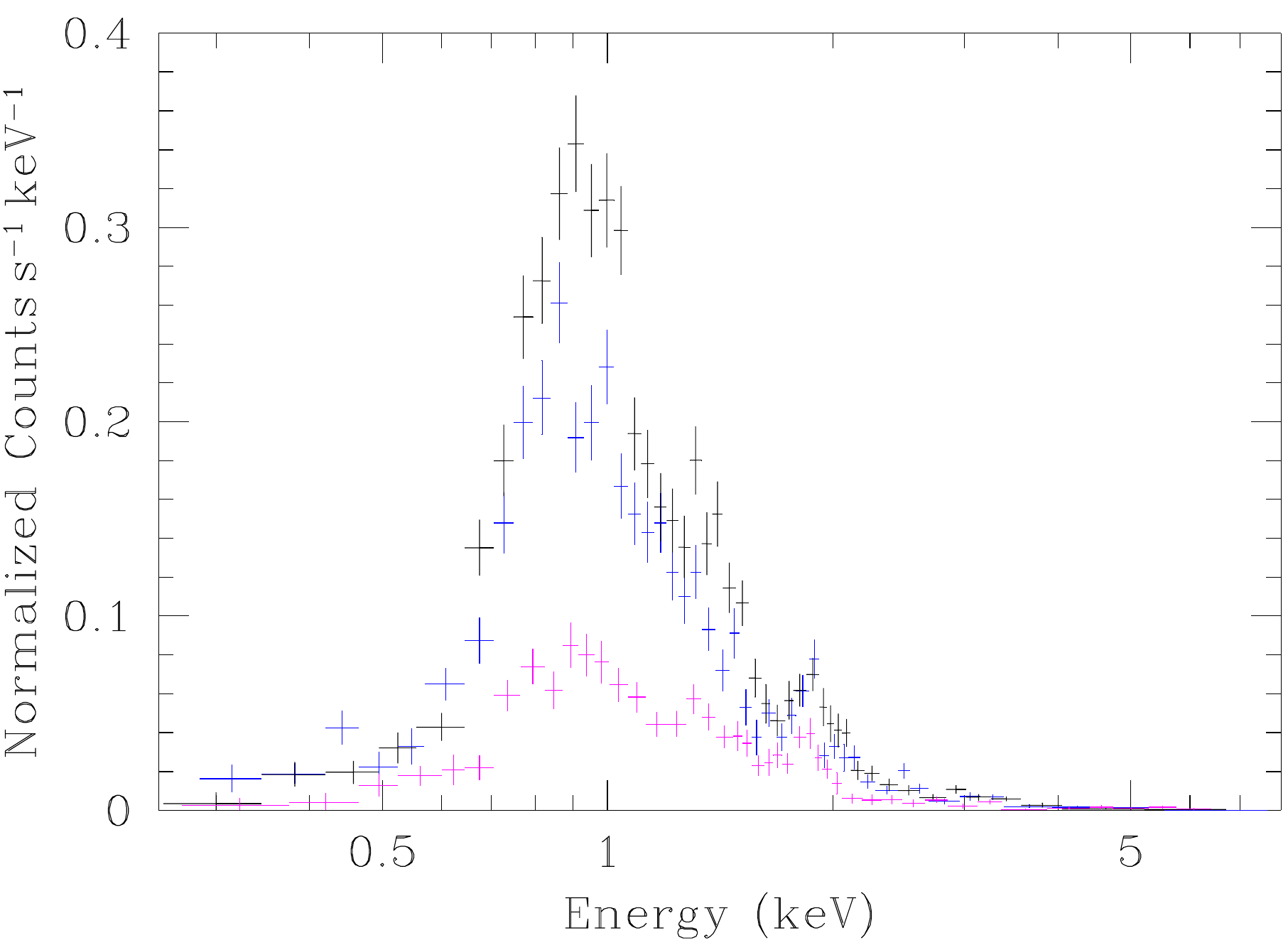}} 
\end{center}
\end{minipage}
\hfill
\begin{minipage}{8cm}
\begin{center}
\resizebox{8cm}{!}{\includegraphics{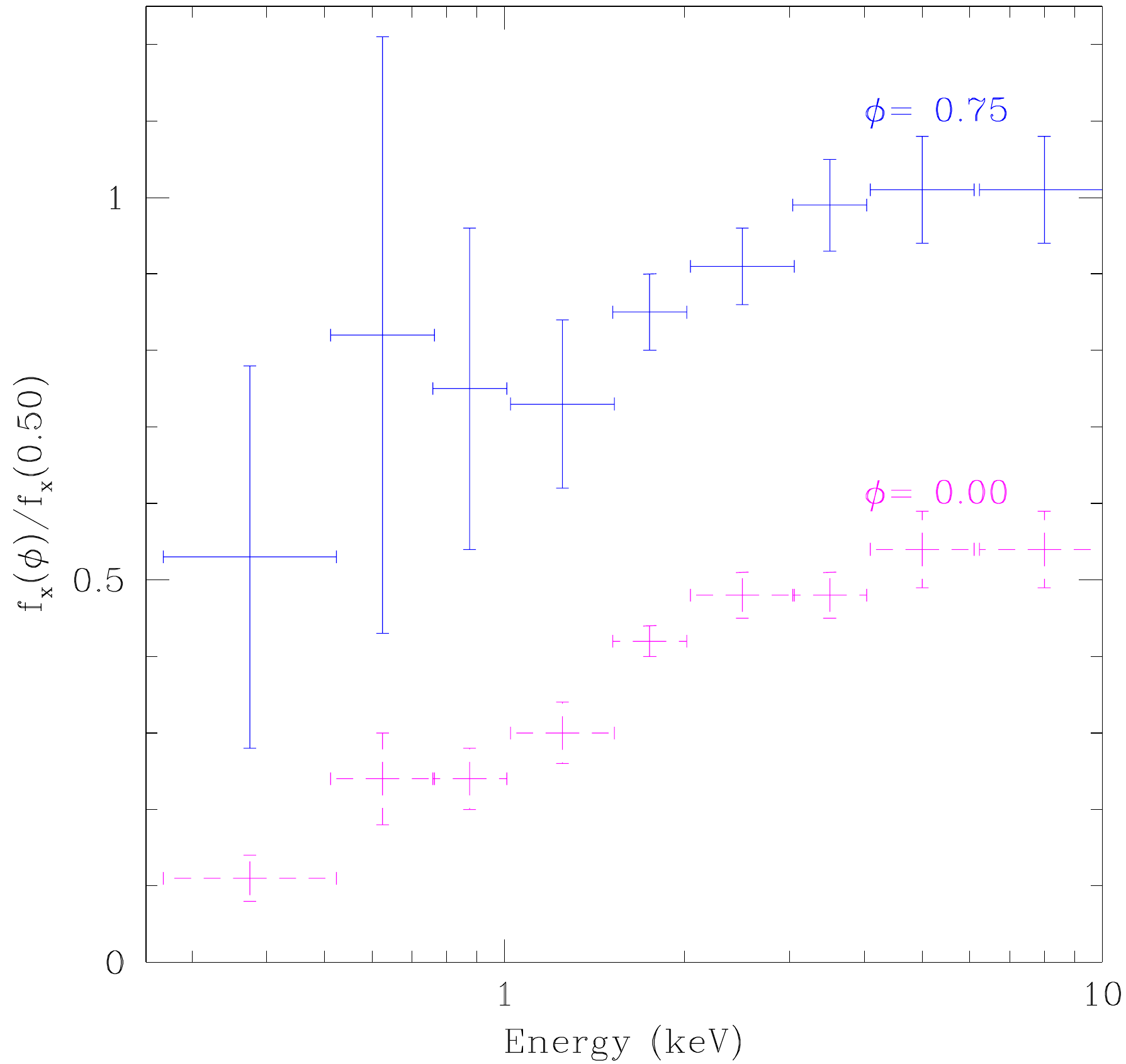}} 
\end{center}
\end{minipage}
\caption{Left: EPIC-pn spectra of HDE\,228766 at the three orbital phases. The black, blue, and magenta symbols stand for phases 0.50, 0.75, and 0.00, respectively. Right: ratio of the observed X-ray fluxes over a series of narrow energy bands with respect to the fluxes at phase 0.5.\label{fx}}
\end{figure*}

The SPM $U\,B\,V$ photometric data were reduced using a multinight, multistar, multifilter method as described by Manfroid (\cite{Manfroid}). This allowed us to take all the measurements of each constant star into account. The data were first cleaned on the basis of the internal rms deviation of the individual integrations. An iterative procedure was then used to reject suspicious values of the
constant stars. Because of the lack of a suitable number of standard star observations, no color transformations were made, besides a zero-point shift based on literature values for the comparison stars. The rms of the deviations of the magnitude difference between the two comparison stars is $0.013$ in $U$ and $0.010$ in $B$ and $V$ (see Table\,\ref{photometry} below). The estimated errors on the photometric data account for the photon noise and for the uncertainties on the changing sky conditions.

To complement our own photometric data, we have retrieved archival photometric data from the Northern Sky Variability Survey (NSVS, Wo\'zniak et al.\ \cite{Wozniak}) and from the Wide Angle Search for Planets (SuperWASP) archive (Butters et al.\ \cite{Butters}). 
NSVS was conducted with the first generation Robotic Optical Transient Search Experiment (ROTSE-I) and provided unfiltered photometry of stars down to a magnitude of 15 and with an angular resolution of 14\,arcsec. There are three entries in the NSVS database that correspond to HDE\,228766: Obj.\,ID 5725284, 8474777, and 8495712. After automatic flag rejection, only the former and the latter remain. These multiple entries are due to overlaps between three fields of the survey. Because of PSF variations across the field, there is a slight difference in the median magnitude of the object entries in NSVS (9.272 for Obj.\,ID 5725284 versus 9.324 for Obj.\,ID 8495712). As we are interested in the relative variations, we combined the data by shifting the light curves of the two object IDs to a zero magnitude baseline (see the forthcoming Fig.\,\ref{NSVS}).
SuperWASP\footnote{\tt http://wasp-planets.net} is an ultra-wide field photometric survey conducted with an array of eight cameras installed on Roque de los Muchachos on La Palma. It provides broadband visible (4000 -- 7000\,\AA) photometry for stars with magnitudes between 7 and 15 and at a plate scale of 13.7\,arcsec/pixel (Butters et al.\ \cite{Butters}). The SuperWASP archive contains photometric data of HDE\,228766 collected over five consecutive nights between 21 and 25 October 2007. 

\section{The X-ray spectra}
The three orbital phases covered by our data correspond to the conjunction phases with the evolved secondary (respectively the primary) star in front for $\phi = 0.00$, (respectively, $\phi = 0.50$) and to the quadrature phase $\phi = 0.75$ with the primary star moving away from the observer. 

Figure\,\ref{fx} illustrates the EPIC-pn spectrum of HDE\,228766 at the three orbital phases. Variations are clearly seen, especially at lower energies. To further quantify these variations, we have used the best-fit spectral models (see Sect.\,\ref{models} below) to evaluate the observed fluxes for each observation over a series of narrow energy bands: 0.25 -- 0.5, 0.5 -- 0.75, 0.75 -- 1.0, 1.0 -- 1.5, 1.5 -- 2.0, 2.0 -- 3.0, 3.0 -- 4.0, 4.0 -- 6.0, and 6.0 -- 10.0\,keV. These fluxes were then divided by those observed at phase 0.5. The results are shown in the right panel of Fig.\,\ref{fx}. 

It is quite obvious that the flux level of the observation at phase 0.0 is lowest in all energy bands, especially in the soft bands. At phase 0.75, the flux is again lower than at phase 0.5 in the lower energy bands but reaches the same level as at phase 0.5 at energies above 3\,keV. These characteristics indicate that the changes in flux level are due to a changing absorption along the line of sight toward the X-ray emitting plasma. In the remainder of this section, we introduce the various ingredients that we have used to fit the spectra. 

\subsection{Absorption by the stellar winds}
The X-ray emission from the wind-wind interaction zone of the binary is seen through the tenuous wind of the primary star at phase $0.50$, whilst we see it through the denser wind of the secondary star at phases $0.75$ and $0.00$. 

Based on the analysis of Rauw et al.\ (\cite{hde}), the primary and secondary stars were classified as O7\,III-I and Of$^+$/WN8ha, respectively. To simulate the absorption by the winds, we computed NLTE model atmospheres with the CMFGEN code (Hillier \& Miller \cite{HM}) based on the typical parameters of an O7\,III-I star (Martins et al.\ \cite{Martins}) and on the parameters of the secondary star inferred by Rauw et al.\ (\cite{hde}). We further computed a third model, where the secondary star has the typical properties of an O4\,If supergiant to account for the alternative classification proposed by Sota et al.\ (\cite{Sota}, see also Sect.\,\ref{intro}). The parameters of the various models are listed in Table\,\ref{cmfgen}. The results of these computations were used to evaluate the wind opacity in the energy range 0.25 -- 12\,keV which is relevant for the EPIC spectra studied in this paper.

\begin{table}
\caption{CMFGEN model parameters. \label{cmfgen}}
\begin{center}
\begin{tabular}{l c c c}
\hline
                 & Model 1    & Model 2      & Model 3 \\
                 & O7\,III-I  & Of$^+$/WN8ha & O4\,If   \\ 
\hline
$T_{\rm eff}$ (K)  & 36000      & 38000        & 40500   \\
$R_*$ (R$_{\odot}$)& 16         & 17           & 19      \\
$log(g)$ (cgs)   & 3.5        & 3.5          & 3.65    \\
$v\,\sin{i}$ (km\,s$^{-1}$) & 200 & 100       & 100     \\
$\dot{M}$ (M$_{\odot}$\,yr$^{-1}$) & $1 \times 10^{-6}$ & $1 \times 10^{-5}$ & $4 \times 10^{-6}$ \\
$v_{\infty}$ (km\,s$^{-1}$) & 3000 & 1800       & 2900    \\
$\beta$          & 1.0        & 1.5          & 0.9     \\
$n_{\rm He}/n_{\rm H}$ & 0.085    & 0.2          & 0.085   \\
$\epsilon_{\rm N}/\epsilon_{\rm N, \odot}$ & 1.0 & 6.0  & 1.0 \\ 
$\epsilon_{\rm N}/\epsilon_{\rm N, \odot}$ & 1.0 & 0.08 & 1.0 \\
$\epsilon_{\rm N}/\epsilon_{\rm N, \odot}$ & 1.0 & 0.08 & 1.0 \\
\hline
\end{tabular}
\end{center}
\end{table}

\begin{figure}[h!tb]
\begin{center}
\resizebox{8cm}{!}{\includegraphics{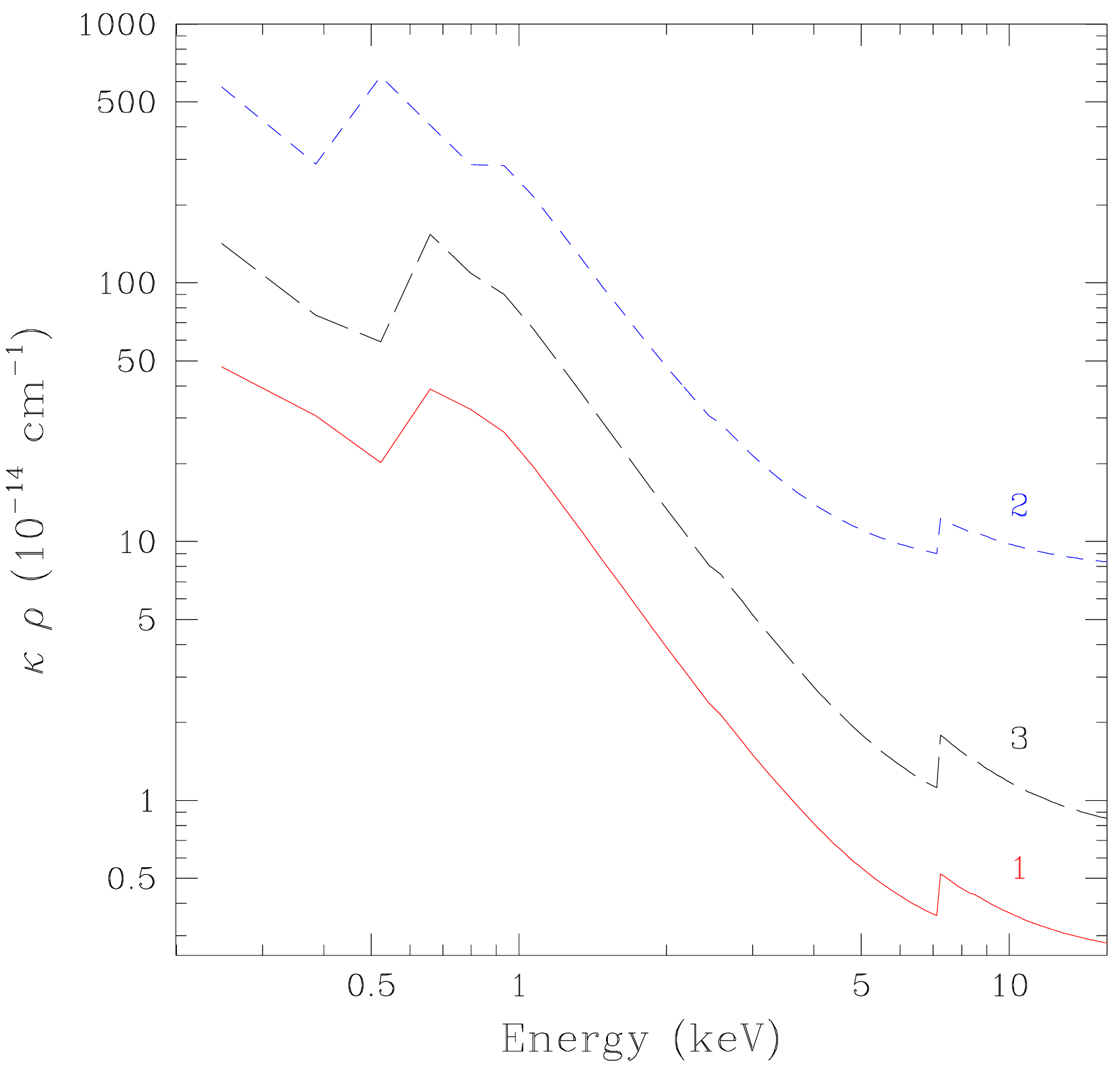}}
\end{center}
\caption{The product of the wind opacity $\kappa(E)$ and the wind density $\rho$ at 2\,$R_*$, as a function of energy for the three model atmospheres. The continuous, short-dashed and long-dashed lines correspond to models 1, 2 and 3, respectively.\label{tau}}
\end{figure}

Figure\,\ref{tau} illustrates the product $\kappa(E)\,\rho$ at a position 2\,$R_*$ from the center of the star for the three models as a function of energy $E$. This quantity is a good proxy of the optical depth corresponding to the different assumptions in Table\,\ref{cmfgen} as $$\tau_E = \int \kappa(E)\,\rho\,ds$$ where the integral is evaluated along a given line of sight. 
Whilst the overall shape of these curves is quite similar above 1\,keV, there are clearly differences around 0.5\,keV which mainly reflect the different compositions of the atmospheres of Model 2 (enhanced nitrogen, depleted carbon and oxygen) compared to Models 1 and 3 (solar composition). Moreover, it is obvious that Model 2 has by far the largest optical depth at all energies.    

In the following, we have used the different model atmosphere calculations to estimate the wind optical depth toward the wind-wind interaction zone, both on theoretical grounds (Sect.\,\ref{toy}) and by spectral fits to the actual data (Sect.\,\ref{models}). For the latter purpose, we have built table models that can be used within {\tt xspec} v.12.7 (Arnaud \cite{Arnaud}) to represent the absorption by the wind column. The opacities used for these table models were evaluated at 2\,$R_*$. In the fitting procedure of the X-ray spectra, the photoelectric absorption by the wind material is expressed in terms of the wind column density $N_{wind}$ through the relation
$$\tau_E = N_{wind}\,\kappa(E, 2\,R_*)\,\overline{m}$$ where $\overline{m}$ is the mean mass of the ions in the wind and $\kappa(E, 2\,R_*)$ is the wind opacity (in units $cm^2\,g^{-1}$) at an energy $E$ and at a distance of $2\,R_*$ from the center of a given star.

\subsection{Spectral fits \label{models}}
We have tested several combinations of optically thin thermal plasma {\tt apec} models (Smith \& Brickhouse \cite{apec}). The absorption by the interstellar medium was computed following the relation between the neutral hydrogen column density $N_H$ and the color excess $E(B-V)$ given by Bohlin et al.\ (\cite{Bohlin}). The $B-V$ colors of HDE\,228766 quoted by Reed (\cite{Reed}) yield $E(B-V) \simeq 0.89$ hence $N_H = 0.52 \times 10^{22}$\,cm$^{-2}$. The additional absorption due to the wind opacity was evaluated from the table models described in the previous section.

It became immediately clear that a good fit of the data could only be achieved if two plasma components with different temperatures were included. At first, we have assumed that both components undergo the same wind absorption and have solar composition according to Anders \& Grevesse (\cite{AG})\footnote{For the CCD spectra investigated here, the revision of the solar composition (see Asplund et al.\ \cite{Asplund}) has little impact on the results of the fit.}. These assumptions yield a plasma temperature of the hotter component that changes widely with the orbital phase. Such a situation is not expected in a colliding wind binary system with a circular orbit. Therefore, as a next step, we have assumed that each plasma component has its own wind column density. This assumption not only results in better quality fits at each phase but leads also to stable values of the plasma temperatures as a function of phase. The cooler and hotter components are found to have $kT_1 = 0.33 \pm 0.05$ and $kT_2 = 0.93 \pm 0.02$\,keV, respectively, where the errors correspond to the dispersions of the best fit values. As a next step, these temperatures were thus frozen in the fits. 

The results of the spectral fits are displayed in Table\,\ref{fits}. Owing to the severe absorption, the spectral parameters of the cooler plasma component, including the wind column density, are less well defined than those of the hotter one. We have considered two different options for the chemical composition of the emitting plasma: solar or non-solar with the He and CNO abundances taken from Rauw et al.\ (\cite{hde}). The latter option leads to somewhat poorer fit qualities, and we thus focus on the results obtained by assuming a solar composition for the emitting plasma. 
\begin{table*}
\begin{center}
\caption{Best fit parameters of the X-ray spectra of HDE\,228766.}
\label{fits}
\begin{tabular}{c c c c c c c c c c c c c}
\hline
Obs & $\phi$ & Model & $\log{N_{wind,1}}$ & $\tau_{wind,1}$ & $10^{4} \times $ norm$_1$ & $\log{N_{wind,2}}$ & $\tau_{wind,2}$ & $10^{4} \times$ norm$_2$ & $\chi^2_{\nu}(\nu)$ & $f_X$ & $f_X^{un}$ \\
\cline{11-12} 
\vspace*{-3mm}\\ 
    &        &       & (cm$^{-2}$) & & (cm$^{-5}$) & (cm$^{-2}$) &      & (cm$^{-5}$) &              &        \multicolumn{2}{c}{$10^{-13}$\,erg\,cm$^{-2}$\,s$^{-1}$} \\
\hline
\vspace*{-3mm}\\
I   & 0.75 & 2 & $21.65^{+.68}$ & 0.11 & $4.2^{+3.5}_{-.9}$ & $22.74^{+.06}_{-.05}$ & 1.38 & $7.19^{+.49}_{-.56}$ &1.75(115) & $3.67^{+.01}_{-.54}$ & $13.4$ \\
\vspace*{-3mm}\\
I   & 0.75 & 3 & $21.23^{+.53}$ & 0.13 & $4.8^{+3.4}_{-1.2}$ & $22.26^{+.07}_{-.06}$ & 1.41 & $6.91^{+.49}_{-.51}$ &1.75(115) & $3.67^{+.00}_{-.45}$ & $13.8$ \\
\vspace*{-3mm}\\
II  & 0.50 & 1 & $21.88^{+.17}_{-.48}$ & 0.50 & $9.2^{+4.4}_{-3.9}$ & $22.16^{+.06}_{-.06}$ & 0.96 & $6.88^{+.60}_{-.62}$ &1.40(119) & $4.55^{+.03}_{-.18}$ & $17.5$ \\
\vspace*{-3mm}\\
III & 0.00 & 2 & $22.53^{+.23}$ & 0.85 & $3.9^{+4.6}_{-2.8}$ & $22.92^{+.11}_{-.09}$ & 2.09 & $4.00^{+.53}_{-.55}$ &1.29(91) & $1.58^{+.02}_{-.14}$ & $4.7$ \\
\vspace*{-3mm}\\
III & 0.00 & 3 & $22.07^{+.26}_{-.50}$ & 0.91 & $5.9^{+11.4}_{-3.8}$ & $22.54^{+.49}_{-.14}$ & 2.69 & $3.97^{+3.03}_{-.50}$ &1.26(91) & $1.59^{+.02}_{-.12}$ & $5.3$ \\
\vspace*{-3mm}\\
\hline
\end{tabular}
\tablefoot{All fits were performed with an {\tt xspec} model of the kind {\tt phabs*(wind$_1$*apec$_1$ + wind$_2$*apec$_2$)}, where {\tt phabs} accounts for the absorption by the interstellar medium with a fixed neutral hydrogen column density of $5.2 \times 10^{21}$\,cm$^{-2}$ and where the {\tt apec} emission components have fixed temperatures of 0.33\,keV (component 1) and 0.93\,keV (component 2) and solar composition. The additional absorptions by the stellar winds are evaluated using the opacities of the model identified in column 3 (see also Table\,\ref{cmfgen}). The wind column densities of the soft plasma component sometimes have large uncertainties and in a few cases, the lower boundary of the uncertainty range could not be defined. The normalization parameter of the {\tt apec} components is defined as $10^{-14}/(4\,\pi\,d^2)\,\int n_e\,n_H\,dV$, where $d$ is the distance of the star and $n_e$ and $n_H$ are the electron and hydrogen number densities of the emitting plasma. The observed fluxes in the 0.5 - 10\,keV energy band and the corresponding fluxes corrected for the ISM absorption are given in the last two columns. The errors on the observed fluxes were obtained via the {\it flux err} command under {\tt xspec}.}
\end{center}
\end{table*} 

The two observations where absorption by the secondary wind is important are treated by adopting the opacities of either Model 2 (Of$^+$/WN8ha parameters) or Model 3 (O4\,If parameters, see Table\,\ref{cmfgen}). The quality of the fit is either identical ($\phi = 0.75$) or slightly better ($\phi = 0.0$) for the opacities from Model 3. The spectra at $\phi = 0.5$, when the primary wind is in front of the wind interaction zone, are analysed by adopting the opacities from Model 1.

The best-fit wind column densities vary from one observation to the other. To compare results obtained with different opacity assumptions, we have converted the wind column densities into optical depths at an energy of 1\,keV. The choice of this energy is motivated by Fig.\,\ref{fx}, which indicates that the X-ray spectral energy distribution of HDE\,228766 peaks around 1\,keV. The results are again listed in Table\,\ref{fits}. The optical depths toward both plasma components are clearly largest when the secondary star is in front ($\phi = 0.0$). 
For the hotter plasma component, which is likely associated with the apex of the wind-collision region, the optical depth is smallest when the primary is in front and more than doubles when the secondary is in front. The cooler plasma component is likely associated with the outer regions of the wind interaction zone, hence undergoing less absorption than the hotter component. The result that the observed wind optical depth of the cooler plasma is lowest at quadrature phase suggests that the opening angle of the wind-wind collision region must be quite large. Indeed, for a sufficiently wide opening angle, the outer parts of the interaction zone that point toward the observer undergo very little wind absorption. 

\section{Multiwavelength light curves \label{lc}}
\subsection{Short-term X-ray variability}
The EPIC light curves of the source and background were extracted using the SAS task {\it epiclccorr}. This tool produces equivalent on-axis, full PSF count rates, providing secure parameters and easing comparison between different exposures. The adopted background and source regions were the same as for the spectral analysis. To avoid very large errors and bad estimates of the count rates, we discarded bins which have an effective exposure time $<$50\% of the time bin length. Our previous experience with {\it XMM-Newton} has shown us that such bins degrade the results. The source light curves that we analyzed were background-corrected.

We choose to extract light curves for three time bins (100\,s, 500\,s, and 1\,ks) and in four energy bands: total (0.2 -- 10.0\,keV), soft (0.3 -- 1.0\,keV), medium
(1.0 -- 3.0\,keV), and hard (3.0 -- 10.0\,keV). The definition of the narrow energy bands was motivated by Fig.\,\ref{fx}, which clearly shows the different impact that wind absorption has on the fluxes in the three bands. Short-term variations in the soft band are most sensitive to variations of the absorption, while short-term variations of the intrinsic emission are more likely to affect the hard band. Unfortunately though, the source count rate in the hard band is close to the background value, thus preventing a meaningful analysis, and we shall not discuss this band further.

We then performed a $\chi^2$ test for several null hypotheses: constancy, linear variation, and quadratic variation. We also compared the improvement of the $\chi^2$ when increasing the number of parameters in the model (e.g., linear trend vs.\ constancy) by means of F-tests. 

The EPIC-MOS data of all three observations are compatible with the constancy hypothesis at the 1\% significance level, and including a linear or quadratic trend does not produce a significant improvement of the $\chi^2$. For the EPIC-pn data, we found that the behavior depends on the way the high-background events are discarded. Applying the most severe rejection of the soft proton flares, we find no significant variation in the EPIC-pn light curves either. 

We thus conclude that there are no significant variations on a timescale of a few hours in either of the X-ray light curves of HDE\,228766. 

\subsection{Constraints from the UV/optical light curve \label{eclipses}}
\begin{figure*}[htb]
\begin{minipage}{6cm}
\begin{center}
\resizebox{6cm}{!}{\includegraphics{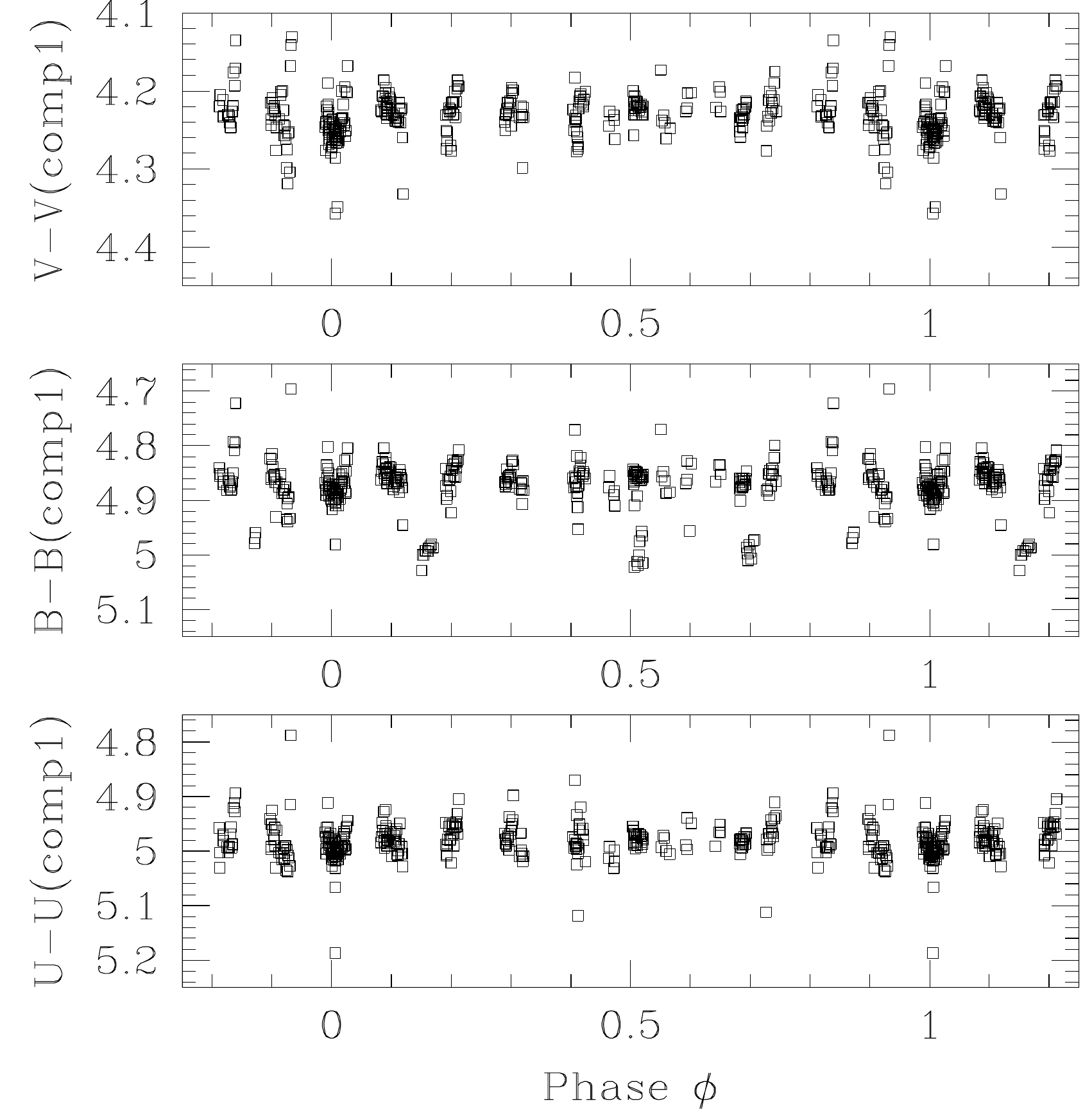}} 
\end{center}
\end{minipage}
\begin{minipage}{6cm}
\begin{center}
\resizebox{6cm}{!}{\includegraphics{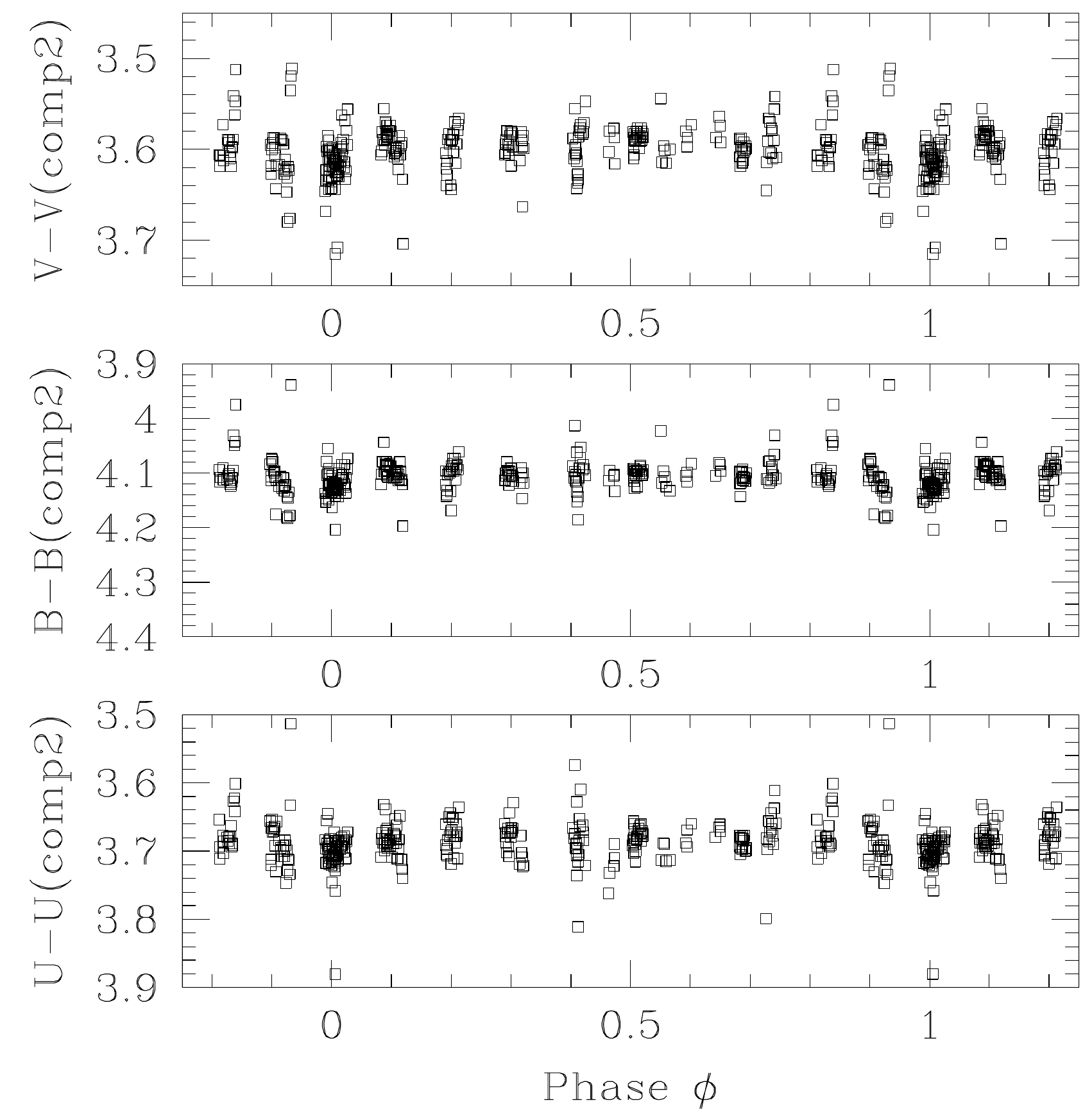}} 
\end{center}
\end{minipage}
\begin{minipage}{6cm}
\begin{center}
\resizebox{6cm}{!}{\includegraphics{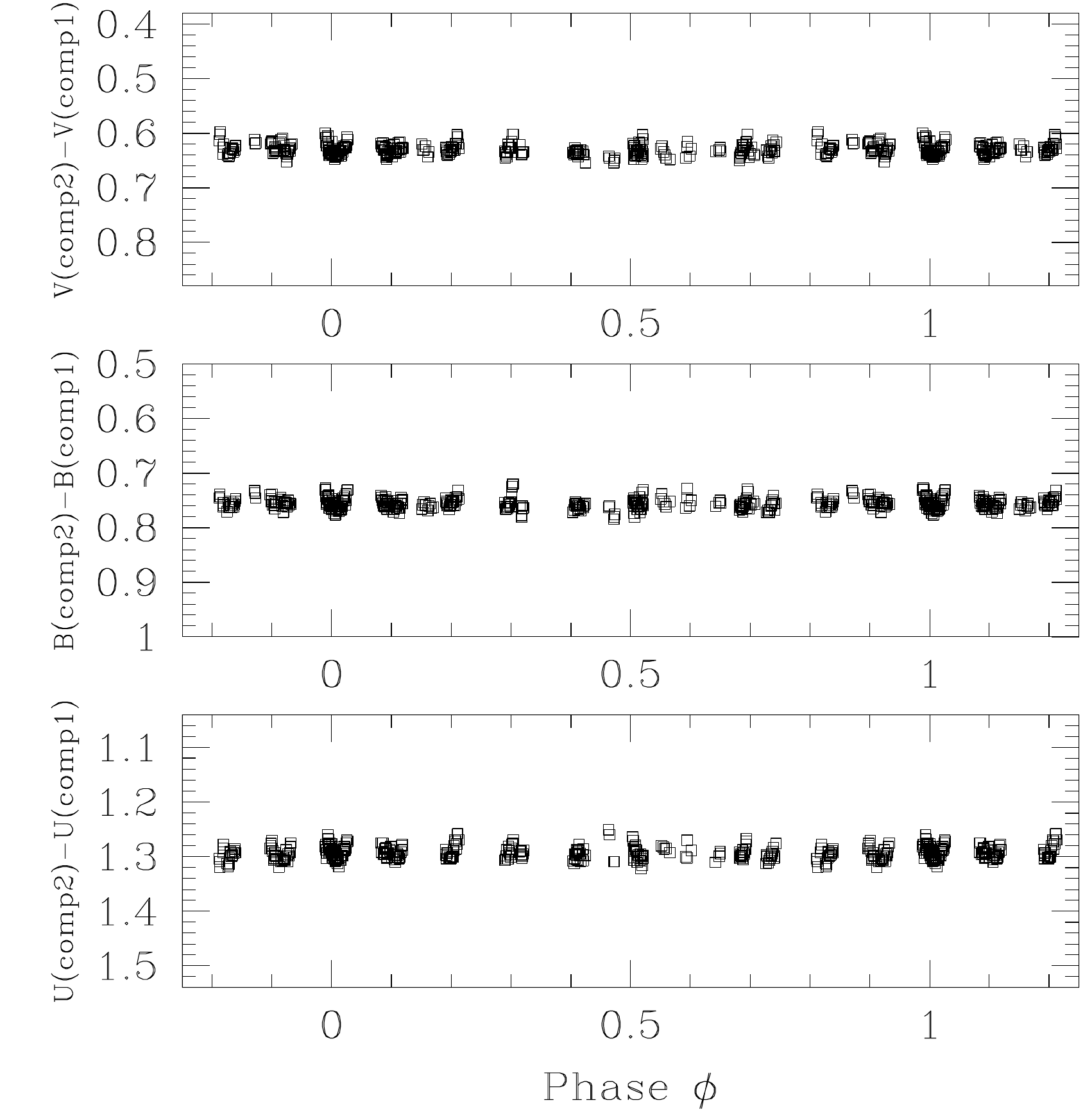}} 
\end{center}
\end{minipage}
\caption{$U\,B\,V$ light curves of HDE\,228766 as a function of orbital phase built from our SPM photometric data. The first and second columns of panels correspond to differential photometry with respect to comparison star HD\,188892 and HD\,193369, respectively. The rightmost column of panels illustrates the difference between the two comparison stars.\label{SPM1}}
\hfill

\begin{minipage}{6cm}
\begin{center}
\resizebox{6cm}{!}{\includegraphics{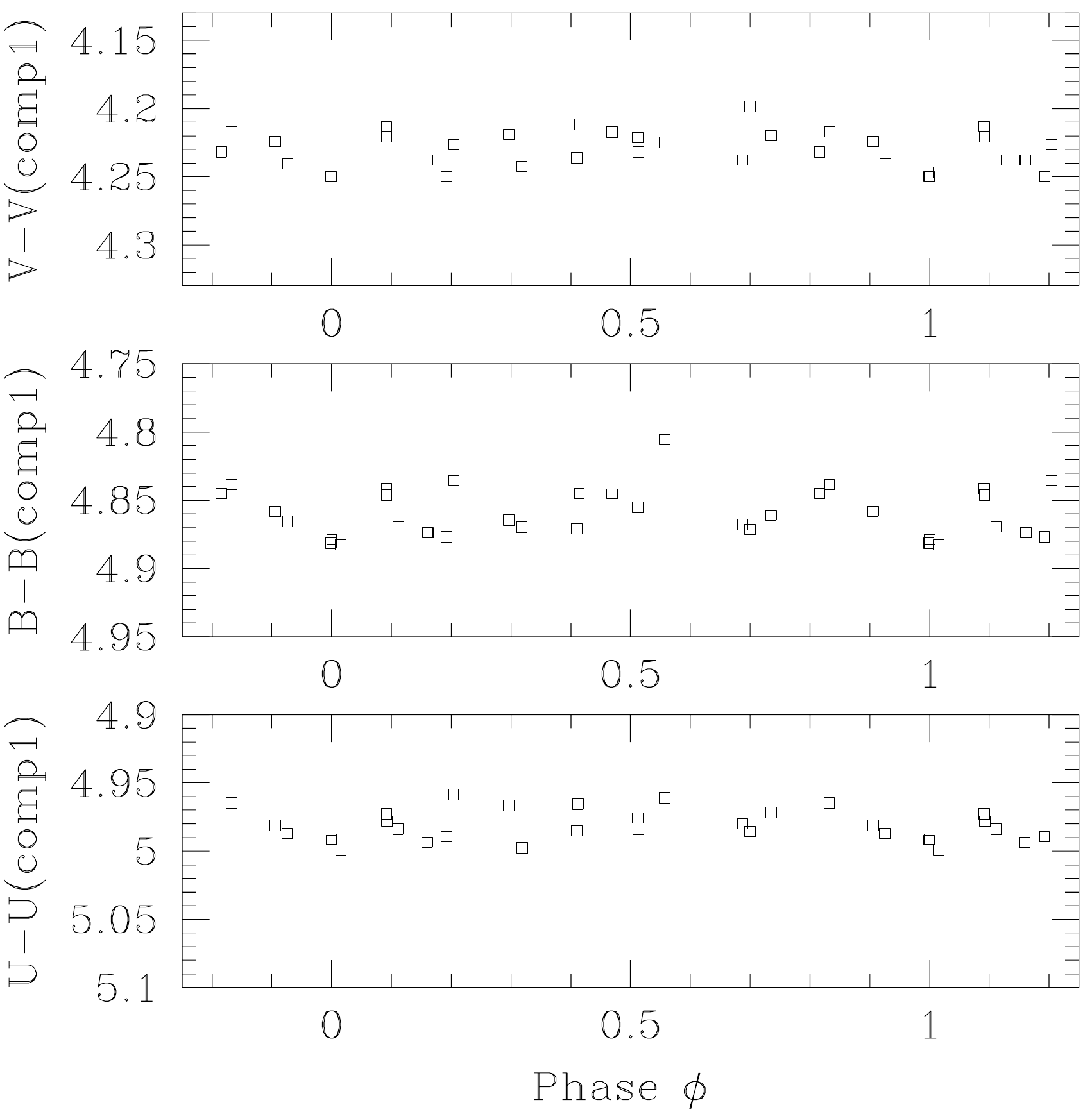}} 
\end{center}
\end{minipage}
\begin{minipage}{6cm}
\begin{center}
\resizebox{6cm}{!}{\includegraphics{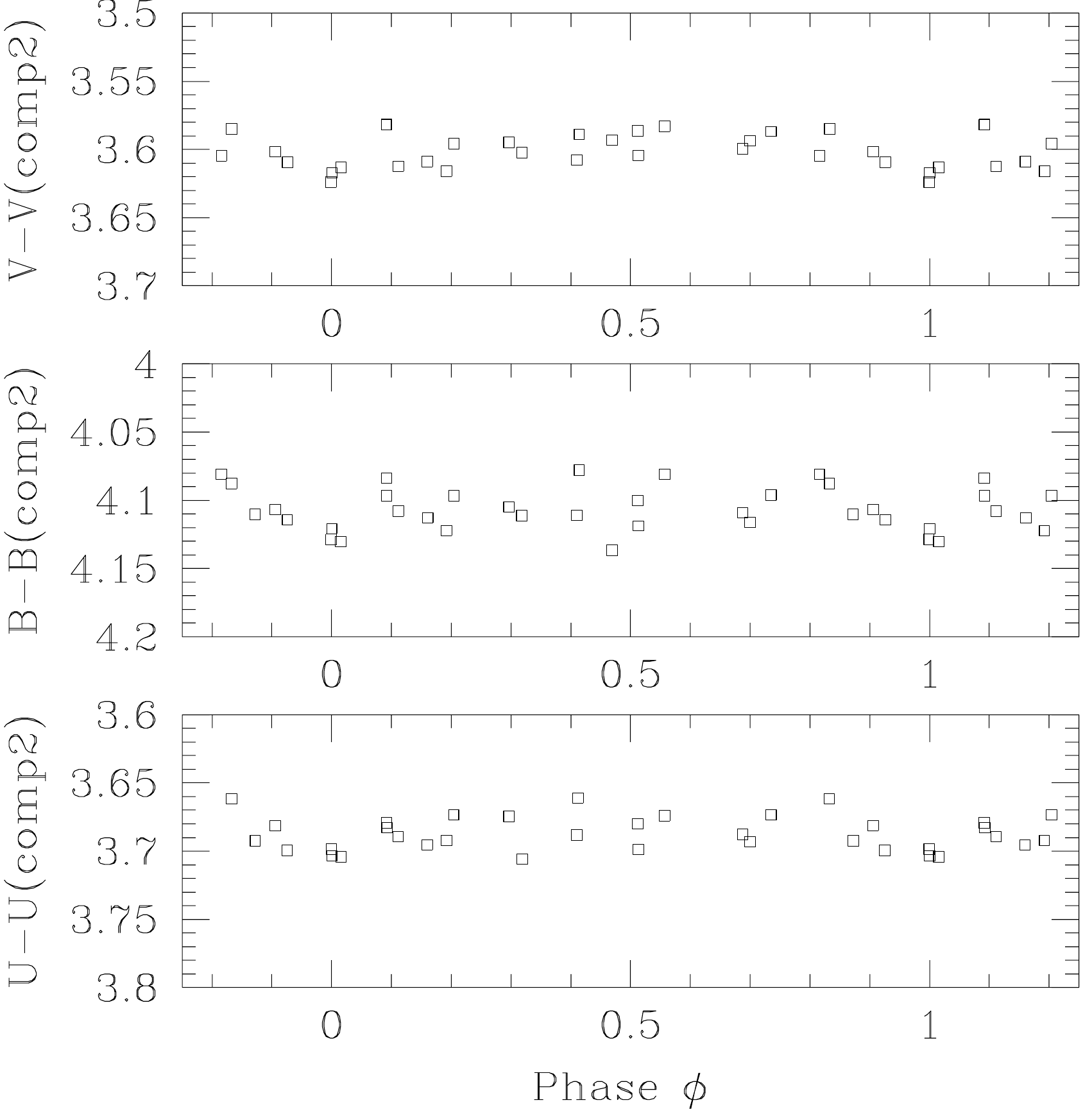}} 
\end{center}
\end{minipage}
\begin{minipage}{6cm}
\begin{center}
\resizebox{6cm}{!}{\includegraphics{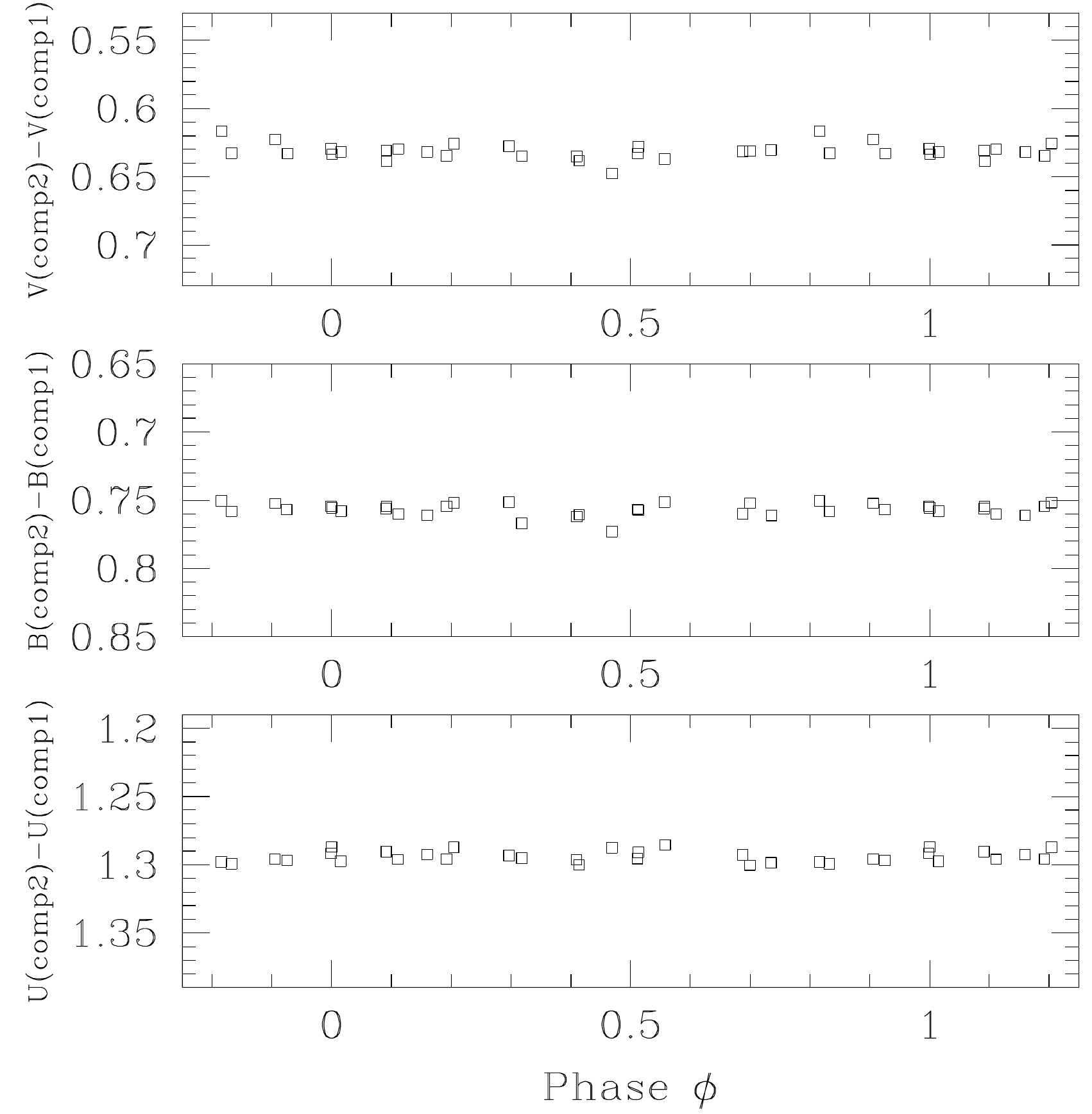}} 
\end{center}
\end{minipage}
\caption{Same as Fig.\,\ref{SPM1}, but after averaging all the data taken during a specific observing night and retaining only those nights where at least 5 observations had been obtained. \label{SPM}}
\end{figure*}

Our SPM $U\,B\,V$ photometry of HDE\,228766 is summarized in Table\,\ref{photometry} and shown in Fig.\,\ref{SPM1}. Table\,\ref{photometry} lists the mean magnitude, the mean error on an individual data point, and the differential photometry with respect to the two comparison stars. The dispersion of the differential photometry is between 1.4 and 1.9 times larger than the typical error on an individual data point. This suggests that HDE228766 might exhibit low-level photometric variability. However, most of this variability occurs on timescales much shorter than the orbital period. We have thus averaged the data taken within a given night for those nights where at least five measurements were taken and discarded data from nights with fewer observations (see Fig.\,\ref{SPM}). The differential light curves corrected in this way have much lower dispersions (0.013\,mag for the $U$ filter and 0.010\,mag in both $B$ and $V$). This confirms our impression that the bulk of the variability, if real, occurs on timescales well below the orbital period. Possible causes of these short-term variations could be pulsations or statistical fluctuations of the number of clumps in the wind of the secondary star.
\begin{table*}
\begin{center}
\caption{Summary of our SPM photometric data in the $U\,B\,V$ filters.\label{photometry}}
\begin{tabular}{c c c c c c c c c c c c c}
\hline
Filter & \multicolumn{2}{c}{HDE\,228766} & Number of points & \multicolumn{3}{c}{HDE\,228766 - comp1} &  \multicolumn{3}{c}{HDE\,228766 - comp2} & \multicolumn{2}{c}{comp1 - comp2}\\
\cline{2-3} \cline{5-7} \cline{8-10} \cline{11-12}
       & Mean magnitude & Mean Error & & $\overline{\Delta\,mag}$ & $\sigma_{\rm raw}$ & $\sigma_{\rm avg}$ & $\overline{\Delta\,mag}$ & $\sigma_{\rm raw}$ & $\sigma_{\rm avg}$ & $\sigma_{\rm raw}$ & $\sigma_{\rm avg}$\\
\hline
$U$ & 9.329 & 0.028 & 331 & 4.982 & 0.032 & 0.012 & 3.689 & 0.030 & 0.014 & 0.013 & 0.004 \\
$B$ & 9.733 & 0.023 & 330 & 4.864 & 0.029 & 0.014 & 4.107 & 0.027 & 0.012 & 0.010 & 0.005 \\
$V$ & 9.177 & 0.021 & 335 & 4.231 & 0.027 & 0.012 & 3.600 & 0.026 & 0.012 & 0.010 & 0.006 \\
\hline
\end{tabular}
\tablefoot{The comparison stars comp1 and comp2 are HD\,188892 (B5\,IV) and HD\,193369 (A2\,V), respectively.}
\end{center}  
\end{table*}
Inspecting the light curves, before (Fig.\,\ref{SPM1}) and after averaging the data taken within a single night (Fig.\,\ref{SPM}), one gets the impression that the system might be marginally fainter around phase 0.0 than at other phases. This could indicate a shallow atmospheric eclipse of the primary star by the denser wind of the secondary, as seen in many WR binary systems (e.g., Gosset et al.\ \cite{EG}, Lamontagne et al.\ \cite{Lamontagne}). However, in view of the averaged light curve, any orbital variation would have a peak-to-peak amplitude of 0.03\,mag at most and certainly less than 0.05\,mag.    

\begin{figure}[htb]
\begin{center}
\resizebox{9cm}{!}{\includegraphics{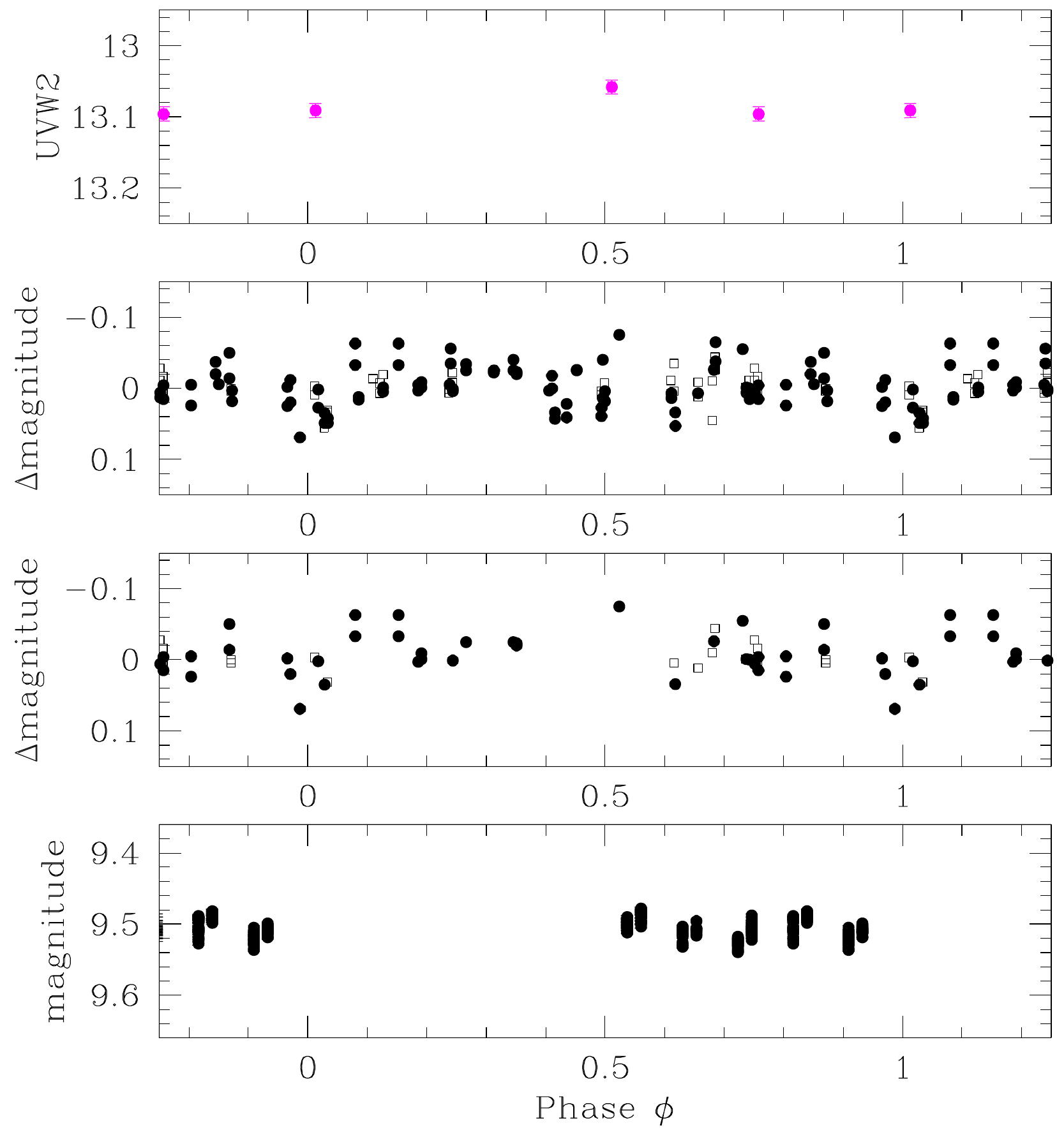}} 
\end{center}
\caption{Top panel: OM light curve of HDE\,228766 in the $UVW2$ filter. Second and third panels from top: NSVS light curve of HDE\,228766. Open squares stand for data from Obj.\,ID 5725284, while filled circles indicate Obj.\,ID 8495712 in the NSVS database (Wo\'zniak et al.\ \cite{Wozniak}). In the second panel, all 117 data points are shown, including those which were obtained via a deblending procedure. The third panel shows only those 42 data points that remain after rejecting the measurements obtained via PSF deblending. Bottom panel: SuperWASP light curve of HDE\,228766.\label{NSVS}}
\end{figure}

The {\it XMM-Newton} OM $UVW2$ measurements of HD\,228766 and the NSVS light curve are shown in Fig.\,\ref{NSVS}. Though we only have three $UVW2$ data points, they support the lack of an eclipse at phase 0.0. Indeed, the star displays the same brightness in the UV at $\phi = 0.75$ and $\phi = 0.0$. 
For the NSVS light curve, the estimated errors on individual data points are 0.011\,mag. When considering all 117 data points, the dispersion about the mean of the light curve amounts to 0.028\,mag, which is 2.5 times the photometric error. When the measurements obtained via deblending (i.e., Flag = 2) are removed, the dispersion remains at 0.029\,mag. We thus conclude that the NSVS data also suggest a moderate level of photometric variability. Folded in orbital phase (see Fig.\,\ref{NSVS}), we note that there are some hints that the system could be fainter around phase 0.0 and possibly also around phase 0.5. However, it seems very much unlikely that there could be eclipses deeper than 0.05 magnitudes.
Unfortunately, the SuperWASP data cover less than half the orbital cycle and miss the critical conjunction phases (see Fig.\,\ref{NSVS}). 

Adopting the stellar parameters from Table\,\ref{cmfgen}, we have used the {\tt NIGHTFALL} code (version 1.70)\footnote{This code was developed and is maintained by Wichmann, Kuster, and Risse, see {\tt \tiny{ http://www.hs.uni-hamburg.de/DE/Ins/Per/Wichmann/Nightfall.html}}} to simulate a set of light curves for different values of the orbital inclination. For inclinations $i \geq 70^{\circ}$, the simulations predict primary and secondary minima deeper than 0.10\,mag. Moreover, the actual eclipses are predicted to last about 0.1 in phase. This behavior is clearly not supported by our data. For an inclination of $68^{\circ}$, the models still predict grazing eclipses in combination with ellipsoidal variations. The resulting minima would have depths of 0.05\,mag. At $i = 65^{\circ}$, only ellipsoidal variations with a peak-to-peak amplitude near 0.02\,mag remain. From these comparisons, we conclude that the photometric data suggest a probable upper limit on the orbital inclination of $68^{\circ}$. 

\section{Modelling the variations of the wind optical depth \label{toy}}
\subsection{The nature of the wind interaction zone \label{nature}}
The gas in the wind interaction zone of a massive binary is either in the radiative or adiabatic regime, depending on the ratio between the timescales of radiative and adiabatic cooling. Stevens et al.\ (\cite{SBP}) showed that the nature of the wind interaction zone can be evaluated via the parameter $$\chi = v_8^4\,\frac{d_{12}}{\dot{M}_{-7}}$$ where $v_8$ is the pre-shock wind velocity in $10^3$\,km\,s$^{-1}$, $d_{12}$ is the separation between the center of the star and the wind interaction zone in $10^{7}$\,km, and $\dot{M}_{-7}$ is the mass loss rate in $10^{-7}$\,M$_{\odot}$\,yr$^{-1}$.

From the parameters in Table\,\ref{cmfgen}, we can infer a theoretical wind momentum ratio of $\eta = \frac{(\dot{M}\,v_{\infty})_2}{(\dot{M}\,v_{\infty})_1}$ between 4 and 6 depending on whether the secondary is an O4\,If or an Of$^+$/WN8ha star. This implies that the stagnation point of the wind-wind collision should be located at 0.67 -- 0.71\,$a$ from the secondary star, where $a$ is the orbital separation of the stars , in the absence of radiative braking (Gayley et al.\,\cite{Gayley}) or radiative inhibition (Stevens \& Pollock \cite{SP}). 

The orbital solution of Rauw et al.\ (\cite{hde}) yields $a\,\sin{i} = 76$\,R$_{\odot}$. Given the rather large minimum masses of the stars, Rauw et al.\ (\cite{hde}) argued that the inclination $i$ could be rather close to $90^{\circ}$. However, in view of our new results (see Sect.\,\ref{eclipses} and the forthcoming Sect.\,\ref{toymodel}), an inclination somewhere between 68 and 55$^{\circ}$ appears more appropriate. One thus expects $a$ to be in the range 82 -- 93\,R$_{\odot}$. In this way, we obtain $\chi \simeq 1400$ -- 1600 for the primary star's wind. For the secondary, we obtain $\chi = 41$ -- 47 or 695 -- 789,  respectively, depending on whether the Of$^+$/WN8ha or the O4\,If classification holds. In either case, we have $\chi >> 1$, which corresponds to an adiabatic regime for the shocked winds of both stars. Parkin \& Sim (\cite{PS2}) recently introduced the concept of self-regulated shocks in colliding wind binaries. These authors showed that the ionization of the material close to the shock by the X-rays emitted in the shock can inhibit the wind acceleration, leading to a lower pre-shock velocity. This effect could thus lower the actual $\chi$ parameter, although the $\chi$ values inferred in our case are so large that even accounting for the self-regulation of the shocks, they are likely to remain in the adiabatic regime. We shall come back to this point in Sect.\,\ref{Discussion}.

Pittard \& Stevens (\cite{PS}) showed that the stronger wind generally dominates the X-ray emission in an adiabatic wind interaction zone, and the dominant factor seems to be the ratio of the $\chi$ parameters of both winds. The smaller $\chi$ corresponds to a more rapidly cooling and, thus, more efficiently radiating wind. In light of the above discussion about the $\chi$ values of HDE\,228766, we would thus expect that the shocked wind of the secondary star accounts for most of the intrinsic X-ray emission of the wind interaction zone. 

\begin{figure}[htb]
\begin{center}
\resizebox{9cm}{!}{\includegraphics{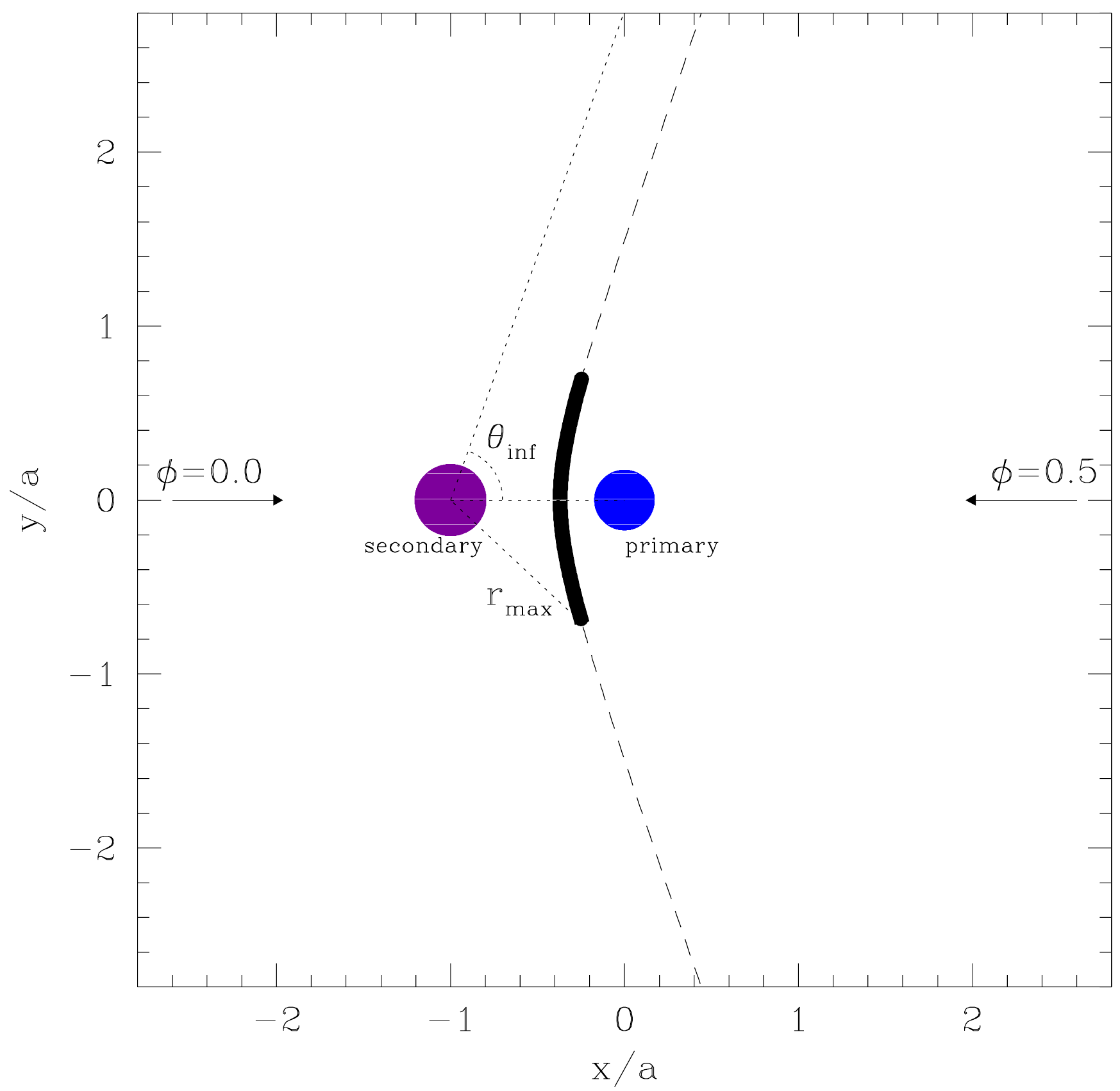}} 
\end{center}
\caption{Schematic view of the geometry of the toy model of the wind interaction zone. The primary star is shown at the origin of the axes, while the secondary is located at $x/a = -1$. The shock front is computed according to the formalism of Cant\'o et al.\ (\cite{Canto}). Only the thick part of the wind intercation zone (corresponding to a maximum radius $r_{\rm max}$ from the center of the secondary contributes to the X-ray emission). The angle $\theta_{\infty}$ corresponds to the asymptotic shock cone opening angle. The arrows yield the projections of the line of sight at the conjunction phases onto the orbital plane.\label{scheme}}
\end{figure}

\subsection{A toy model of the wind interaction zone \label{toymodel}}
To interpret the results obtained in the previous sections, we build a toy model of the wind-wind interaction region based on the formalism of Cant\'o et al.\ (\cite{Canto}) for non-accelerated hypersonic flows that collide in a thin interaction zone contained between two hydrodynamical shocks. For a given set of parameters (mass-loss rate and wind velocity of each star, stellar radii, and orbital separation), the shape of the contact discontinuity and the surface density $\sigma$ of the shocked material are computed, respectively, according to equations (23) and (30) of Cant\'o et al.\ (\cite{Canto}). The widths of the wind interaction zone is then computed as $\sigma/\rho_{\rm post}$ where $\rho_{\rm post}$ is the post-shock density, which amounts to $4 \times \rho_{\rm wind} = \frac{\dot{M}}{\pi\,r^2\,v(r)}$ for an adiabatic shock. This model yields the geometry of the wind interaction zone. For a given orbital inclination and orbital phase, we then evaluate the optical depth due to the wind material along the line of sight from each cell of material in the wind interaction zone to the observer. In the conventional $(p,z)$ coordinates of the star that is in front, the optical depth of a cell at coordinates $(p,z)$ is expressed as
$$\tau_E (p,z) = \frac{2\,\tau_{E,*}}{\sqrt{\frac{p^2}{R_*^2}-1}}\,\left(\frac{\pi}{2} - \arctan{\frac{\frac{\alpha\,p}{R_*} - 1}{\sqrt{\frac{p^2}{R_*^2}-1}}}\right)$$
where $\alpha = \tan{(\frac{\pi - \theta}{2})}$, $\theta = \arccos{\frac{p}{\sqrt{p^2 + z^2}}}$, and $\tau_{E,*} = \frac{\kappa(E, 2\,R_*)\,\dot{M}}{4\,\pi\,v_{\infty}\,R_*}$.
We further search for intersections between the line of sight and the stellar bodies and discard those cells that are occulted from the flux calculations. 

The intrinsic X-ray emission by the plasma in the wind interaction region is assumed to scale with $\rho_{\rm post}^2\,Q(E,T)$ where $Q(E,T)$ is the emissivity of the plasma at temperature $T$ and energy $E$. The plasma temperature is computed by following the Rankine-Hugoniot shock jump condition. The emissivities are either taken as constant over the temperature range of the shock-heated plasma, or assumed to behave as the emissivity of the Ne\,{\sc x} Ly$\alpha$ line at 1.022\,keV. This assumption turns out to have no impact on the flux ratios computed with our model. We further consider different hypotheses on the extent of the X-ray emitting zone. This is described via the parameter $r_{\rm max}$, which yields the outer limit of the emitting region, as measured from the center of the primary star (see Fig.\,\ref{scheme}). We test values of $r_{\rm max}$ equal to 0.25, 0.5, 0.75, 1.0, 1.5, 2.0, 2.5, 3.0, 4.0, and 100 times the orbital separation.
  
In short-period massive binaries, the Coriolis force leads to a deflection of the shock cone. This effect is not included in our simple model based on the Cant\'o et al.\ (\cite{Canto}) formalism. However, as the X-ray emission is dominated by the region around the apex of the shock cone, we do not have to consider the full complexity of the spiral geometry of the outer parts of the wind interaction zone. Thus, as a first approximation, we can estimate the deflection angle at the apex of the cone as $$\delta = \arctan{\frac{v_{\rm orb}}{v_{\rm wind}}}$$ (Marchenko et al.\ \cite{Marchenko}, Stevens \& Howarth \cite{SH}, Gosset et al.\ \cite{Gosset}). In the case of HDE\,228766, the orbital solution of Rauw et al.\ (\cite{hde}) yields $v_{\rm orb} \simeq 360$\,km\,s$^{-1}$. Neglecting the effect of radiative braking, inhibition, and self-regulation, we then estimate deflection angles between $7$ and $11^{\circ}$. These values are rather small and are thus neglected in our simple model.

We can then use our code to simulate observed fluxes at $E = 1$\,keV and to estimate the flux-weighted optical depth toward the wind interaction zone at the same energy. We do so for three orbital phases ($\phi = 0.0$, 0.5 and 0.75).
The opacities of the primary wind are assumed to correspond to those of an O7\,III-I star (see Model 1 in Table\,\ref{cmfgen}), while we consider the two alternative assumptions (Models 2 and 3 in Table\,\ref{cmfgen}) for the opacities of the secondary wind. These simulations are performed over a grid of 35721 models sampling five parameters: an orbital inclination between 50 and 90$^{\circ}$ (with a step of $5^{\circ}$), $\dot{M}_1$ between $0.2$ and $2.2 \times 10^{-6}$\,M$_{\odot}$\,yr$^{-1}$ in steps of $0.2 \times 10^{-6}$\,M$_{\odot}$\,yr$^{-1}$, $v_{\infty,1}$ and $v_{\infty,2}$ each between 1500 and 3000\,km\,s$^{-1}$ by steps of 250\,km\,s$^{-1}$, and a wind-momentum ratio $\eta$ between 1.5 and 21.5 by steps of 3. The latter sampling determines the secondary star's mass loss rate via $\dot{M}_2 = \eta\,(\dot{M}_1\,v_{\infty,1})/v_{\infty,2}$. 

The results of our simulations are then compared to the observed flux ratios (in the 0.9 -- 1.1\,keV range) $f_X(\phi=0.0)/f_X(\phi=0.5) = 0.23^{+.01}_{-.04}$ and $f_X(\phi=0.75)/f_X(\phi=0.5) =0.69^{+.01}_{-.18}$ and to the best-fit optical depth values at the three phases quoted in Table\,\ref{fits}. The ability of the simulation to reproduce the observations is expressed in terms of a $\chi^2$. 

\begin{figure*}[htb]
\begin{center}
\resizebox{18cm}{!}{\includegraphics{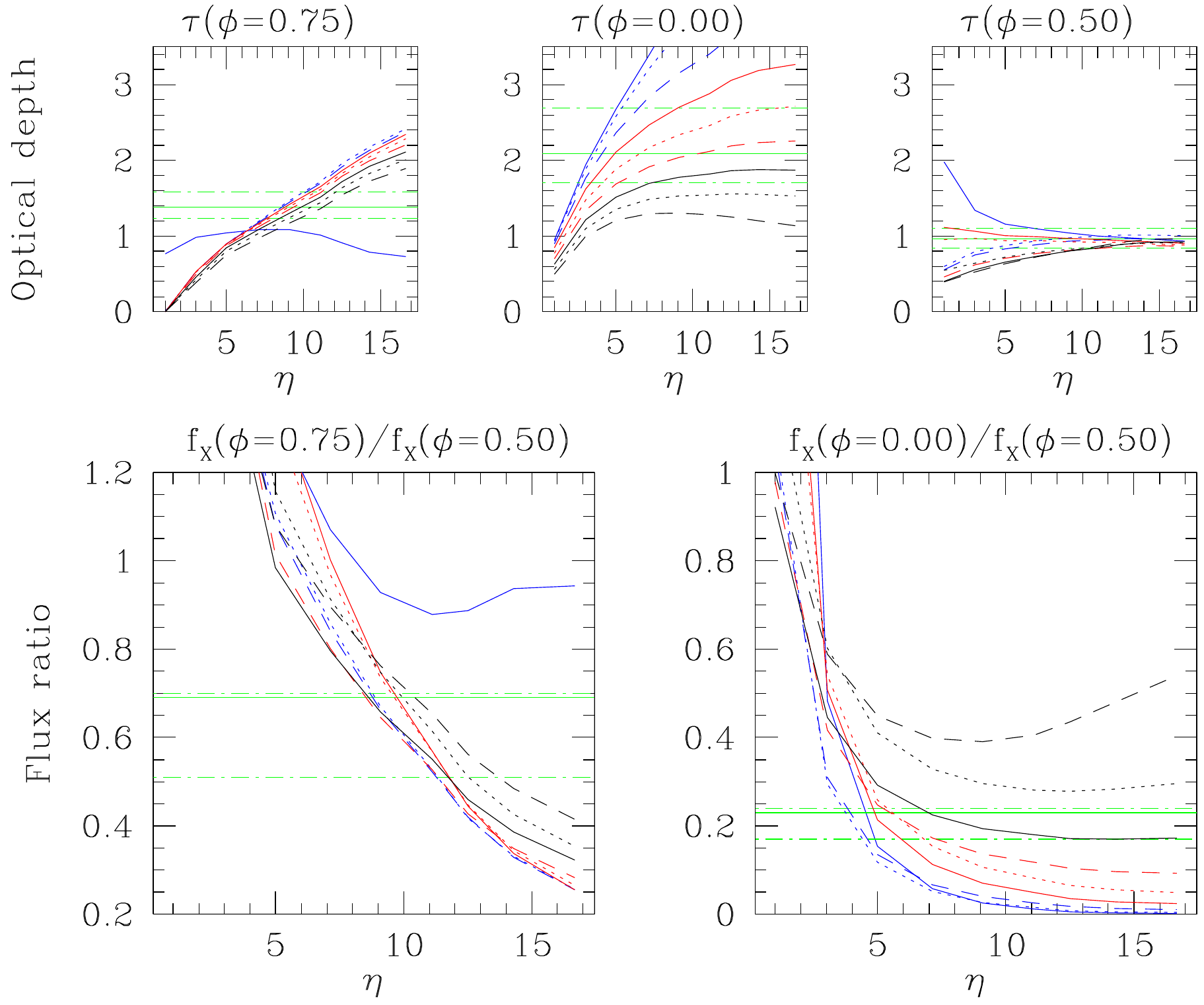}} 
\end{center}
\caption{Output from the toy model for the three optical depths (top panels) and the two flux ratios (bottom panels). The green horizontal lines yield the observed values and their error bars. The blue lines correspond to inclinations of 90, 85 and 80$^{\circ}$ for the solid, dotted, and dashed lines, respectively. The solid, dotted, and dashed red and black lines yield in turn the results for inclinations of 75, 70, 65$^{\circ}$, and 60, 55, 50$^{\circ}$, respectively. The model adopted here uses the opacities from Model 2 and assumes $\dot{M}_1 = 1.87 \times 10^{-6}$\,M$_{\odot}$\,yr$^{-1}$, $v_{\infty, 1} = 2245$\,km\,s$^{-1}$, $r_{\rm max} = 0.75\,a$. \label{toyoutput}}
\end{figure*}

Considering the different assumptions on $r_{\rm max}$, we immediately find that models, where the X-ray emission zone extends well beyond the orbital separation ($r_{\rm max} \geq 4.0\,a$), yield large $\chi^2$ values and are, thus, unable to account for the observed behavior. The same is true for very compact emission zones with $r_{\rm max} \leq 0.25\,a$. In the latter case, the emission zone is completely occulted at both conjunction phases for higher inclination angles. Figure\,\ref{toyoutput} illustrates the output of a specific toy model (with $\dot{M}_1 = 1.87 \times 10^{-6}$\,M$_{\odot}$\,yr$^{-1}$, $v_{\infty, 1} = 2245$\,km\,s$^{-1}$, and $r_{\rm max} = 0.75\,a$) for different inclinations and values of $\eta$. One clearly sees that the optical depth at the $\phi = 0.0$ and $\phi = 0.75$ depends both on the inclination and the shock opening angle (via the value of $\eta$). For $\phi = 0.5$, $\tau$ is less sensitive to $\eta$ and $i$, since we are seeing the shock cone through the less dense primary wind at this phase. 

\begin{table*}
\begin{center}
\caption{Properties of the best solutions obtained from a grid of models as a function of $r_{\rm max}$.\label{toyresults}}
\begin{tabular}{c c c c c c c c c}
\hline
\multicolumn{9}{c}{Model 2 opacities}\\
\hline
$r_{\rm max}$ &  $i$  &  $\dot{M}_1$ & $\dot{M}_2$ & $v_{\infty,1}$ & $v_{\infty,2}$ &$\theta_{\infty}$ & $\chi^2_{\rm min}$ & Number of sol.\\
($a$) & ($^{\circ}$) & ($10^{-6}$\,M$_{\odot}$\,yr$^{-1}$) & ($10^{-6}$\,M$_{\odot}$\,yr$^{-1}$) & (km\,s$^{-1}$) & (km\,s$^{-1}$) & ($^{\circ}$) & & \\
\hline
0.50 & $54.0 \pm 3.7$ & $1.59 \pm 0.46$ & $20.1 \pm 4.6$ & $2380 \pm 479$ & $2032 \pm 397$ & $50 \pm 4$ & 1.19 & 421 \\
0.75 & $57.1 \pm 3.4$ & $1.87 \pm 0.32$ & $22.2 \pm 5.1$ & $2245 \pm 483$ & $2012 \pm 402$ & $51 \pm 6$ &  0.97 & 425 \\
1.0 &  $61.2 \pm 3.2$ & $2.03 \pm 0.21$ & $21.9 \pm 4.9$ & $2046 \pm 444$ & $1903 \pm 342$ & $52 \pm 6$ &  3.07 & 280 \\
1.5 & $74.3 \pm 6.4$ & $2.04 \pm 0.20$ & $20.3 \pm 4.0$ & $2105 \pm 481$ & $1714 \pm 233$ & $54 \pm 5$ &  3.41 & 241 \\
2.0 &  $76.4 \pm 4.4$ & $2.07 \pm 0.17$ & $24.8 \pm 4.7$ & $1944 \pm 435$ & $1733 \pm 242$ & $50 \pm 5$ &  3.99 & 171 \\  
2.5 &  $77.3 \pm 4.2$ & $2.13 \pm 0.12$ & $27.4 \pm 3.9$ & $1692 \pm 261$ & $1674 \pm 178$ & $48 \pm 3$ &  5.00 &  56 \\
3.0 & $76.7 \pm 4.1$ & $2.17 \pm 0.08$ & $30.0 \pm 3.2$ & $1542 \pm 102$ & $1542 \pm 102$ & $47 \pm 2$ & 6.33  &  6 \\
\hline
\multicolumn{9}{c}{Model 3 opacities}\\
\hline
$r_{\rm max}$ &  $i$  &  $\dot{M}_1$ & $\dot{M}_2$ & $v_{\infty,1}$ & $v_{\infty,2}$ &$\theta_{\infty}$ & $\chi^2_{\rm min}$ & Number of sol.\\
($a$) & ($^{\circ}$) & ($10^{-6}$\,M$_{\odot}$\,yr$^{-1}$) & ($10^{-6}$\,M$_{\odot}$\,yr$^{-1}$) & (km\,s$^{-1}$) & (km\,s$^{-1}$) & ($^{\circ}$) & & \\
\hline
0.50 & $53.5 \pm 3.7$ & $1.01 \pm 0.38$ & $6.7 \pm 1.3$ & $2039 \pm 452$ & $2546 \pm 421$ & $53 \pm 5$ &  1.26 & 245 \\
0.75 & $54.6 \pm 3.7$ & $1.55 \pm 0.41$ & $7.6 \pm 1.4$ & $1915 \pm 414$ & $2612 \pm 369$ & $57 \pm 5$ &  0.87 & 298 \\
1.0 & $58.3 \pm 4.5$ & $1.76 \pm 0.37$ &  $7.3 \pm 1.5$ & $2030 \pm 479$ & $2352 \pm 477$ & $62 \pm 7$ &  2.11 & 448 \\
1.5 & $72.6 \pm 8.1$ & $1.82 \pm 0.36$ & $7.1 \pm 1.5$ & $2085 \pm 485$ & $2227 \pm 462$ & $65 \pm 6$ &  1.13 & 682 \\
2.0 & $76.1 \pm 5.8$ & $1.90 \pm 0.31$ & $8.0 \pm 1.7$ & $2016 \pm 476$ & $2207 \pm 469$ & $63 \pm 7$ &  1.19 & 458 \\
2.5 & $77.3 \pm 4.8$ & $2.02 \pm 0.22$ &  $9.0 \pm 1.8$ & $1874 \pm 446$ & $2191 \pm 502$ & $62 \pm 8$ & 3.45 & 166 \\
3.0 & $80.4 \pm 4.5$ & $2.17 \pm 0.07$ & $11.0 \pm 1.9$ & $1522 \pm  72$ & $2533 \pm 428$ & $55 \pm 6$ &  5.96 &  23 \\
\hline
\end{tabular}
\tablefoot{The last two columns yield the lowest value of $\chi^2$ for each value of $r_{\rm max}$ and the number of parameter combinations that have $\chi^2 \leq \min_{r_{\rm max}}(\chi^2_{\min}) + 7.04$, which is the number of solutions within a 68.27\% confidence level for a six parameters problem.}
\end{center}
\end{table*}

We have used our extensive grid of toy models to search for the combination of parameters that best reproduces the five observables. The results are summarized in Table\,\ref{toyresults}. For each value of $r_{\rm max}$, we have computed the mean and standard deviation of the parameters for those models that fall within the 68.27\% confidence level. These are the numbers (along with their errors) given in columns 2 -- 7 of Table\,\ref{toyresults}. 

We clearly see from this table that the best-fit inclination increases with $r_{\rm max}$. This is easily understood if we recall that both a lower inclination and a larger emission region imply a reduction of the optical depth to the  emission region at $\phi = 0.0$. Therefore, to compensate for an increase in $r_{\rm max}$, one must increase $i$. We see that models with $r_{\rm max} \geq 1.5\,a$ imply inclinations that exceed the upper limit that is inferred from the optical light curve of HDE\,228766: our acceptable models then yield inclinations in the range 54 -- $61^{\circ}$. Such inclinations imply actual masses that are between 1.49 and 1.89 times larger than the minimum masses inferred by Rauw et al.\ (\cite{hde}). 

Another conclusion is that fitting the flux ratios and the optical depths implies models with rather high values of $\eta \geq 5$. These models have shock-cone half-opening angles ($\theta_{\infty}$) in the range 47 -- $65^{\circ}$. The corresponding value of the secondary's mass-loss rate depends of course on the assumed spectral-type (as the opacity at 1\,keV strongly depends on the chemical composition of the wind), but we always find values around $2.1 \times 10^{-5}$\,M$_{\odot}$\,yr$^{-1}$ for Model 2\footnote{This value of $\dot{M}_2$ is a factor 2 larger than that inferred by Rauw et al.\ (\cite{hde}), which is from a CMFGEN model atmosphere fit of the optical spectrum.} and $0.7 \times 10^{-5}$\,M$_{\odot}$\,yr$^{-1}$ for Model 3 for a given composition. 

The result that acceptable models have $r_{\rm max}$ in the range 0.5 -- $1.5\,a$ implies that plasma at larger distances from the stars is probably too cool to emit copious X-rays. A priori, adiabatic expansion is unlikely to be sufficient to cool the plasma over such a short distance. The post-shock plasma temperature could also decrease as a result of shock obliquity. However, this will probably only play a role for small shock-cone opening angles below about $40^{\circ}$. Our models yield $\theta_{\infty}$ in the range of 47 -- $65^{\circ}$, which makes this explanation unlikely. Alternatively, the compactness of the emitting region  could be due to radiative cooling. In Sect.\,\ref{nature}, we concluded that the wind interaction zone in HDE\,228766 should be in the adiabatic regime. This result, as derived with the wind parameters of Table\,\ref{cmfgen}, remains valid even if we adopt the larger mass-loss rates derived hereabove. The only way to change this conclusion, would be to assume that the wind velocities are considerably reduced prior to entering the wind interaction zone, possibly as a result of the X-ray ionization of the wind (see Parkin \& Sim \cite{PS2}). Dedicated hydrodynamical simulations of HDE\,228766 are needed to address this point. Such simulations are beyond the scope of the present paper. 

\section{Summary and conclusions \label{Discussion}}
In this paper, we have used new X-ray and optical data to scan the winds of the very massive binary system HDE\,228766 and further constrain its orbital configuration. The optical data revealed no eclipses. The X-ray data in turn reveal a strong phase-dependence of the absorption by the wind material. The considerable X-ray optical depth of the secondary wind likely produces a significant Auger ionization, which explains the oddities of the optical spectrum of this star, especially regarding the simultaneous presence of N\,{\sc iii}, N\,{\sc iv}, and N\,{\sc v} lines. 

Our main conclusions are as follows:
\begin{itemize}
\item[$\bullet$] The orbital inclination is lower than $68^{\circ}$ and probably in the range $54$ -- $61^{\circ}$. The latter interval implies masses for the primary between 47 and 60\,M$_{\odot}$ and for the secondary between 38 and 48\,M$_{\odot}$. While these values are large, they are not unrealistic in view of other very massive systems that have been discovered over the last decade. 
\item[$\bullet$] The bulk of the colliding-wind X-ray emission of HDE\,228766 arises from the innermost part of the wind-wind interaction zone with a maximum radius from the secondary star of $r_{\rm max} \simeq 0.5$ -- $1.5\,a$. 
\end{itemize}
Further insight into the nature of HDE\,228766 will come from an in-depth analysis of its optical spectrum using a spectral disentangling method that we will perform in the near future. This will provide a definite answer regarding the spectral types of the components and will help us assess their chemical composition, and hence, the  evolutionary stage of the system. 
In the longer run, HDE\,228766 appears as a promissing target for Doppler tomography of its X-ray emission lines (see Sciortino et al.\ \cite{Athena}) by using the high-resolution spectrometer on ESA's next-generation X-ray observatory. This study may allow us to map the wind interaction zone in velocity space and possibly reveal the temperature stratification of this wind interaction region.

\acknowledgement{This research was supported by a bilateral convention between Conacyt (Mexico) and FRS-FNRS (Belgium). The Li\`ege team further acknowledges support from Belspo through an XMM PRODEX contract, from the FRS-FNRS, as well as from an ARC grant for Concerted Research Actions, financed by the Federation Wallonia-Brussels. This publication makes use of data from the NSVS survey created jointly by the Los Alamos National Laboratory and University of Michigan. The NSVS was funded by the Department of Energy, NASA and the NSF. We are grateful to Dr P.\ Wo\'zniak for advices in the use of the NSVS data. This paper makes also use of data from the first public release of the WASP data as provided by the WASP consortium and services at the NASA Exoplanet Archive. We thank an anonymous referee for carefully reading our manuscript.}

\end{document}